\documentclass[preprint]{emulateapj}
\usepackage{epsfig}
\journalinfo{To appear in AJ, January 2009}

\newcommand{\etal}{{\frenchspacing\it et al. }}

\newcommand{\lsim}{\hbox{ \rlap{\raise 0.425ex\hbox{$<$}}\lower 0.65ex\hbox{$\sim$} }}
\newcommand{\gsim}{\hbox{ \rlap{\raise 0.425ex\hbox{$>$}}\lower 0.65ex\hbox{$\sim$} }}

\def\plot3#1#2#3{\centering \leavevmode
     \epsfxsize=.3\columnwidth \epsfbox{#1} \hfil
     \epsfxsize=.3\columnwidth \epsfbox{#2} \hfil
     \epsfxsize=.3\columnwidth \epsfbox{#3} \hfil}

\shorttitle{Northern Sky Optical Cluster Survey III.}
\shortauthors{Gal \etal}

\begin{document}

\title{The Northern Sky Optical Cluster Survey III: \\
A Cluster Catalog Covering Pi Steradians}

\author{R.R. Gal, P.A.A. Lopes\altaffilmark{1,4}, R.R. de Carvalho\altaffilmark{1}, J.\ L. Kohl-Moreira\altaffilmark{2}, H.V. Capelato\altaffilmark{1} \& S.G. Djorgovski\altaffilmark{3}}

\affil{Institute for Astronomy, 2680 Woodlawn Dr., Honolulu, HI 96822 \\ 
{\indent Email: rgal@ifa.hawaii.edu} }

\altaffiltext{1}{Instituto Nacional de Pesquisas Espaciais - Divis\~{a}o de Astrof\'{i}sica (CEA), Avenida dos Astronautas 1758, S\~{a}o Jos\'{e} dos Campos, SP 12227-010, Brazil} 

\altaffiltext{2}{Observatorio Nacional/MCT, COAA, Brazil}

\altaffiltext{3}{Astronomy, MS 105-24, Caltech, Pasadena, CA 91125, USA}

\altaffiltext{4}{Current address: IP\&D, Universidade do Vale do Para\'{i}ba, Av. Shishima Hifumi 2911, S. J. dos Campos 12244-000, SP, Brazil}

\begin{abstract}
We present the complete galaxy cluster catalog from the Northern Sky
Optical Cluster Survey, a new, objectively defined catalog of
candidate galaxy clusters at $z\lsim0.25$ drawn from the Digitized
Second Palomar Observatory Sky Survey (DPOSS). The data presented here
cover the Southern Galactic Cap, as well as the less-well calibrated
regions of the Northern Galactic Cap. In addition, due to improvements
in our cluster finder and measurement methods, we provide an updated
catalog for the well-calibrated Northern Galactic Cap region
previously published in Paper II. The complete survey covers 11,411
square degrees, with over 15,000 candidate clusters. We discuss
improved photometric redshifts, richnesses and optical luminosities
which are provided for each cluster. A variety of substructure
measures are computed for a subset of over 11,000 clusters. We also
discuss the derivation of dynamical radii $r_{200}$ and its relation
to cluster richness. A number of consistency checks between the three
areas of the survey are also presented, demonstrating the homogeneity
of the catalog over disjoint sky areas. We perform extensive
comparisons to existing optically and X-ray selected cluster catalogs,
and derive new X-ray luminosities and temperatures for a subset of our
clusters. We find that the optical and X-ray luminosities are well
correlated, even using relatively shallow ROSAT All Sky Survey and
DPOSS data. This survey provides a good comparison sample to the
MaxBCG catalog based on Sloan Digital Sky Survey Data, and complements
that survey at low redshifts $0.07<z<0.1$.

%In addition, we discuss systematic uncertainties
%in richness measurements, and their impact on the use of richness as a
%proxy for cluster mass.
\end{abstract}


\section{Introduction}
The construction of large catalogs of galaxy clusters for use in
studies of cosmology, large scale structure, and galaxy evolution has
often proven to be a difficult task \citep[see][for a
review]{gal08}. Indeed, the last such catalog generated using optical data,
and covering the entire high-galactic-latitude Northern sky was that
of \citet{abe58}, updated in 1989 \citep{aco89}. More recently, the
Northern ROSAT All-Sky (NORAS) Galaxy Cluster Survey \citep{boh00} has
provided an X-ray selected catalog covering a similar region, but with
many fewer clusters, while the largest catalog using purely digital
observations \citep{koe07b} covers a smaller area, albeit with much
better photometry.

Because much improved optical data has become available, with
automated techniques to generate objective, well-characterized cluster
samples, we undertook the generation of a modern, optically selected
cluster catalog, the Northern Sky Optical Cluster Survey
\citep[NoSOCS][hereafter Papers I and II]{gal00a,gal03}, and its
deeper extension \citep[hereafter Paper IV]{lop04}. The need for a
modern cluster survey covering a significant portion of the sky is
striking. The catalog from Paper II has already been used to suggest a
connection between short-duration gamma-ray bursts and clusters
\citep{geh05,blo06}, to search for giant arcs \citep{hen08}, to
associate compact groups and large-scale structure
\citep{dec05,and06}, and to examine X-ray and optical cluster
properties \citep{lop06}.  Here we present the second and final
installment of this catalog, including photometric redshifts,
richnesses, optical luminosities and substructure measures. Although
superior imaging data is now available from the Sloan Digital Sky
Survey (SDSS, \citealt{yor00}), the imaged area covers less than half
of the Northern sky, and is not expected to ever reach the area
coverage of our catalog. The currently published cluster catalog from
SDSS \citep{koe07b} covers $\sim7500$ deg$^2$, containing nearly
14,000 clusters. That survey, using the MaxBCG technique, has a lower
redshift cutoff of $z=0.1$; our survey extends down to $z=0.07$. This
provides a sample of more local clusters whose properties can be
examined in detail \citep{lop08}, especially using the extensive SDSS
spectroscopic database.

\begin{figure*}
%\epsscale{0.9}
\plotone{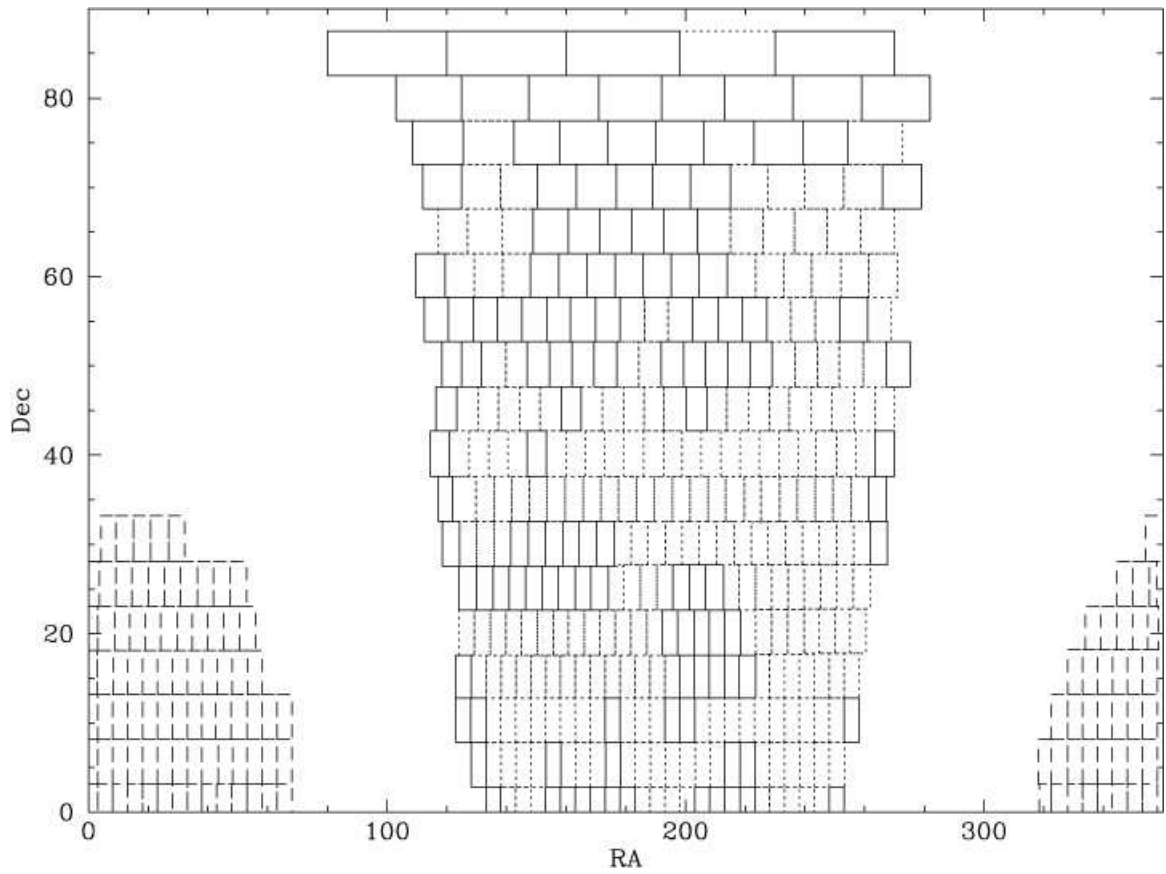}
\caption{The distribution of DPOSS plates used in NoSOCS, showing the NGP well-calibrated (solid lines), poorly calibrated (dotted lines) and SGP (dashed lines) regions.\label{plates}}
\end{figure*}

\begin{figure*}
%\epsscale{0.9}
\plotone{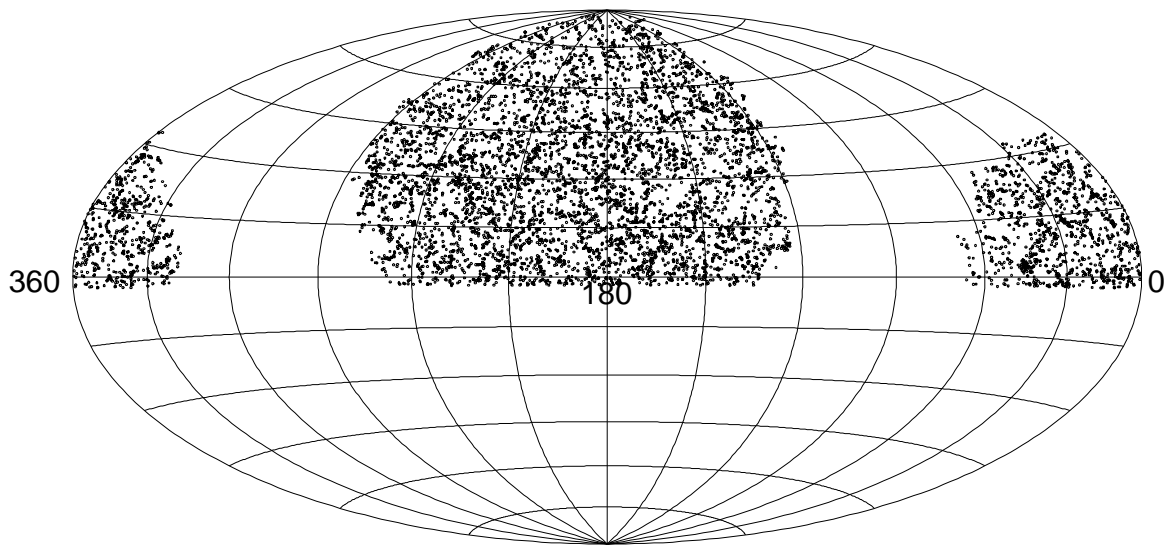}
\caption{Aitoff projection of the complete NoSOCS cluster catalog in equatorial coordinates.\label{alleq}}
\end{figure*}

\begin{figure*}
%\epsscale{0.9}
\plotone{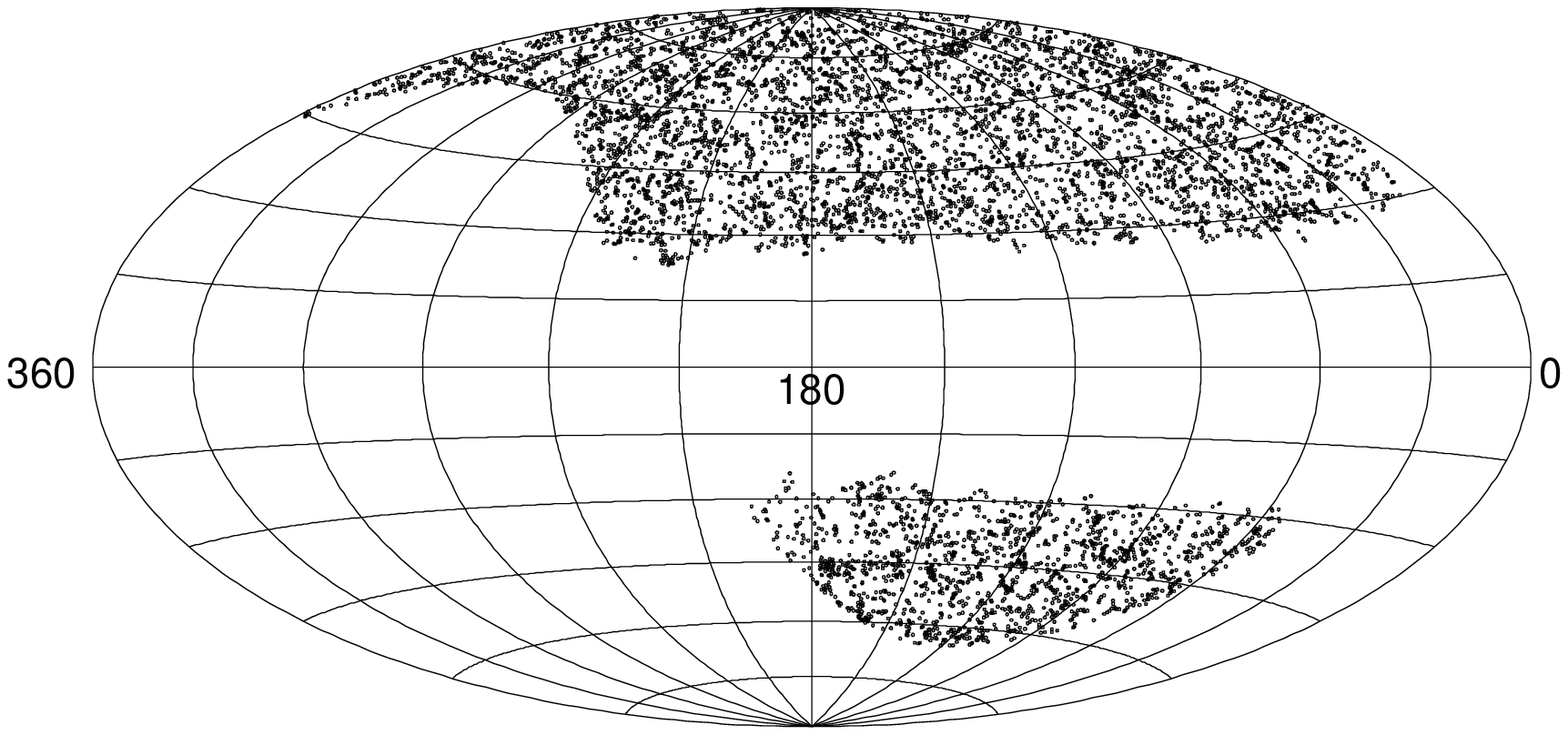}
\caption{Aitoff projection of the complete NoSOCS cluster catalog in galactic coordinates.\label{allgal}}
\end{figure*}

The regions covered in this paper are shown in
Figure~\ref{plates}. Three separate areas are covered: (1) The
well-calibrated North Galactic Pole (NGP) region already described in
Paper II (dotted lines); (2), the portion of the NGP with less well
calibrated plates, not covered in Paper II (solid lines); and (3), the
Southern Galactic Pole region (SGP, dashed lines). The area covered by
the NGP-poor region is 2813$\Box^{\circ}$, while the SGP covers
2917$\Box^{\circ}$. Together with the 5681$\Box^{\circ}$ surveyed in
Paper II, the final NoSOCS catalog covers 11411 square degrees. The
distribution of clusters in the survey is shown in Figures~\ref{alleq}
and~\ref{allgal}, in equatorial and galactic coordinates,
respectively. Only clusters with richness $N_{gals}>20$ are shown for
clarity.

The survey methodology is described in \S2. Although the overall
detection technique is very similar to that discussed in Paper II, we
have modified our definition of bad areas due to very bright objects,
reducing contamination by spurious detections. Changes to the
photometric redshift estimation yield more robust (and realistic)
errors. The richness estimator has also been improved, providing
robust error estimates, as well as total $r$-band optical
luminosities. Then, in \S3, we discuss two new sets of parameters
computed for our cluster sample: estimates of cluster substructure and
the dynamical radius $r_{200}$.

In \S4 we describe the general characteristics of our cluster sample,
and present consistency tests for the three sky regions utilized. The
selection functions describing the completeness as a function of
richness and redshift are presented, as is an estimate of the
contamination by projection effects. Our complete and final cluster
catalog, including an updated version covering the area from Paper II,
is presented. This cluster sample is compared to other
optically-selected catalogs in \S5. In particular, we compare our
catalog to the SDSS MaxBCG catalog of \citet{koe07b}, the only modern optical cluster catalog
covering a similar area and redshift range. We then examine the
correlation between our redshifts and richnesses and the X-ray
measurements from the Northern ROSAT All-Sky Galaxy Cluster Survey
(NORAS, \citealt{boh00}). Because NORAS consists of many fewer
clusters than our catalog, we use our optical positions to measure
X-ray fluxes and upper limits from RASS for a significant subsample of
NoSOCS. 

\section{Survey Methodology}

The detection of galaxy clusters in modern optical imaging surveys
typically utilizes the existence of a tight color--magnitude relation
for cluster galaxies, noted nearly half a century ago
\citep{bau59,bow92}. Surveys at low and high redshift have made use of
this observation to efficiently detect clusters, while achieving low
contamination (false positive) rates \citep{gla00,han05}. We use the
galaxy catalogs from the Digitized Second Palomar Observatory Sky
Survey \citep[DPOSS,][]{djo99} as the basis for our
survey. Unfortunately, the limited photometric accuracy
($\sigma_{mag}\sim0.25^m$ at $r=19.5$, \citealt{gal04}) of DPOSS
forces us to rely solely on the two-dimensional projected galaxy
distribution for cluster detection.  Details of the photometric
calibration and star/galaxy separation are discussed in \citet{gal04}
and \citet{ode04} respectively. In brief, the positions of the galaxies are
used to generate adaptive kernel (AK) density maps \citep{sil86} which
outputs images in units of projected galaxy density.  We then run SExtractor \citep{ber96} on
these images, detecting peaks which are identified as potential galaxy
clusters. We refer the reader to Papers I and II for more
comprehensive descriptions of the cluster detection. Photometric
redshifts are estimated using the background-corrected mean $r$
magnitude and median $g-r$ color of the galaxies within a
0.5$h^{-1}$Mpc radius of the cluster center; details on improvements to this estimator from our previous work are described below. The photometric redshifts are used to recenter the clusters, also discussed later. Richnesses are computed
by counting galaxies with $M^*-1\le M^*_r \le M^*+2$ within the same
radius, with corrections applied for higher (and now lower) redshift
clusters where the faint (or bright) end of this magnitude range is
beyond our catalog limits. We note that galaxy colors are used only in the
post-detection steps to estimate photometric redshifts for the clusters.

\subsection{Enhanced Removal of Bad Areas}

In the process of generating the catalog presented in Paper II, it
became obvious that the DPOSS catalogs were insufficient for finding
very bright stars and galaxies, which are typically deblended into
numerous fainter components.  In later work, Paper IV
utilized the Tycho-2 catalog to excise candidate higher-redshift
clusters in the area of bright stars after detection. To avoid
spurious cluster detections due to these artifacts, we now rely on the
Tycho-2 and RC3 catalogs to exclude regions in the vicinity of bright
objects {\em before} performing cluster detection. Specifically,
around bright stars we exclude circular regions whose area depends on
the star's Tycho magnitude; $2'$ radius for $m_{Tycho}<7.0$, $1.5'$
for $7.0\le m_{Tycho}<8.0$, and $1.0'$ for $8.0\le
m_{Tycho}<9.5$. These radii were chosen by visually inspecting plate
images and the resulting galaxy catalogs to determine the sizes of regions
contaminated by the bright stars. In addition, larger regions around
bright galaxies in the RC3 catalog \citep{dev91} were excised,
corresponding to $5\times r_{\rm{RC3}}$ for $r_{\rm{RC3}}<25''$ and
$8\times r_{\rm{RC3}}$ for $r_{\rm{RC3}}\ge25''$.

%% Should we provide list of bad areas?

The removal of objects in these contaminated regions results in empty
holes in the galaxy catalogs. An undesired consequence is that in
these areas the adaptive kernel artificially increases the smoothing
radius (which is inversely proportional to the local density). To
remedy this problem, we generate simulated galaxy catalogs using the
Raleigh-Levi distribution \citep[Paper II]{pos02}, and use these to
fill the excised areas as well as the densitometry spot region on each
plate. With this technique, the regions which would otherwise be empty
instead contain galaxies with the average projected density for each
plate. 

After cluster detection is completed, the final catalog is again
checked against the list of bad areas. For some very
extended, nearby galaxies, or very bright stars, we found through
visual inspection that the exclusion area described above was not
always sufficient. We therefore increased the radius of these areas,
by a factor of 1.5 for the Tycho-2 stars, and to $10\times r_{\rm{RC3}}$
for the RC3 galaxies. In addition, we created additional bad areas
from catalogs of Galactic globular and open clusters, as these objects
were found to be contaminants in our previous catalogs. Any candidates
found within an exclusion radius defined by the sizes of these
Galactic clusters was also eliminated. Visual inspection of candidates
flagged in this last step shows that the vast majority were indeed
bad. A total of 404 candidates, 2.5\% of the sample, are
removed from the catalog in this step.

%Well, the other test that I did was to compare the density within 0.2
%h^-1 Mpc (dens1) and 0.5 h^-1 Mpc (dens2). If (1.5*dens1) < dens2 I
%select the cluster as a bad candidate. There are 401 clusters in this
%case (list_of_badcls_check_dens.dat). If I simply tried to test dens1
%< dens2, then the number of clusters is way too high (~1800). That's
%why I used (1.5*dens1) < dens2.

\subsection{Photometric Redshift Improvements}

A number of small but significant changes were made to the algorithm
presented in Paper II. These modifications enhance the photometric
redshift measurements based on DPOSS photometry and provide more
robust error estimates. First, we re-derived the empirical relation
between the median $g-r$ color, mean $r$ magnitude of the cluster
galaxies, and the spectroscopic redshift (equation 1 in Paper
II). This was prompted in part by modifications to the background
estimator (described below), which results in changes to the global
properties of the cluster galaxy populations. Our larger sky coverage
also allowed us to restrict the spectroscopic cluster sample to those
clusters with more than three concordant redshifts in the compilation
of \citet{stru91}, resulting in a training sample of 254 clusters. We
also found that the redshift estimator was more reliable when
restricted to galaxies with $m_r\le19.5$, as opposed to $m_r\le20.0$
used before.  The recalibrated photometric redshift relation used here
is
\begin{equation}
{z_{phot} = 0.3273\times(g-r)_{med} - 0.0702\times r_{mean} - 1.2685 }
\end{equation}
with a $Q_{\sigma}(z_{spect} - z_{phot})/(1+z_{spec})=0.023$, an improvement of $\sim30\%$ over the results in Paper II.

As noted above, one major modification to our technique involves
measurement of the fore- and background galaxy contamination in the
cluster area. In Papers I and II, we used color and magnitude
distributions from each DPOSS plate ($\sim30$ deg$^2$), scaled to
the area of each cluster on that plate, as the background
correction. This ignores the contribution of local large scale
structure to the background of each cluster, which can introduce
systematic errors since galaxy colors and luminosities are strongly
correlated with local density \citep{dre80,bla05}. We therefore implemented a local
background estimator, as follows:
\begin{enumerate}
\item A random position is chosen within a background annulus of width $1.3^{\circ}$ starting 3$h^{-1}$ Mpc from the cluster center. 
\item A box of size $20'\times 20'$ is placed at the random location
\item A check is performed to see if the box intersects a bad area (hole) in the survey; if so, we return to step 1.
\item The distribution of colors and magnitudes is generated for the galaxies in this box.
\item The procedure is iterated until ten background regions are successfully measured.
\item The $3\sigma$-clipped medians of the distributions from the ten background regions are used as the background correction for that cluster.
\end{enumerate}

The redshift estimator is run ten times for each cluster
candidate. Changes in the placement of the randomly located background
measurement regions result in variations of the background galaxy
color and magnitude distributions, which effect the final photometric
redshift. By repeating the measurement, we derive an estimate of the
random error in $z_{phot}$ due to the background correction, which we
then add in quadrature to the scatter from the redshift --
photometric properties relation. Although some of the latter is likely
due to difficulties with properly estimating the background
contribution, we prefer to estimate the redshift errors
conservatively, adding the errors as if they were independent. We note
that such problems are likely to be significantly reduced in modern
digital imaging surveys,where the photometric errors are an order of
magnitude smaller than for DPOSS \citep{gal03,ade07}.  As a final
change to the redshift estimator, we allow more iterations for
convergence (15 instead of 10). This was found to increase the number
of successful redshift estimates, while allowing further iteration
simply grew the computational requirements with little improvement.

For some clusters, the photometric redshift estimator does not
converge in all ten of the runs. The final photometric redshift is
taken as the mean $z_{phot}$ from the $n_{successful}$ runs. The
redshifts and their associated errors are provided in Columns 5 and 6
of Table~\ref{allclusters}. For those clusters where the $z_{phot}$
estimate always failed, these three columns are all set to 0. Such
clusters are likely to be spurious detections or have contaminated
photometry from bright stars, telescope reflections or other
artifacts.

\subsection{Cluster Centroids}

During the photometric redshift measurement process, the cluster
positions are also recomputed. At each iteration of the $z_{phot}$
computation, we calculate the median position of the galaxies within a
1$h^{-1}$ Mpc radius of the previously determined center, and this is
taken as the new cluster centroid for the next iteration of the
photometric redshift estimation. To avoid large offsets, the maximum
change in position is limited to 2 arcminutes; if the recomputed
center is further from the previous location, then the center is not
moved. The above steps are repeated for each of the $n_{successful}$
photometric redshifts, and the final cluster position is recorded as
the mean of the corresponding $n_{successful}$ positions. In those
cases where the photometric redshift estimator does not converge, we
retain the original position (from running SExtractor on the density
map).

\subsection{Richness and Luminosity Measures}

Richnesses and luminosities are computed using the basic methods
described in \citet{lop06}, with some further refinements, and
adjusted for the lower redshift range probed here. The procedure
consists of five steps for each cluster, described below.

\begin{enumerate}
\item We use $z_{phot}$ to determine the apparent magnitude $m^*_r$, the aperture corresponding to 0.50 $h^{-1}$ Mpc, and the $k$-corrections $k_e$ and $k_s$ for
elliptical and late-type galaxies (Sbc) at the cluster redshift. We
select all galaxies within 0.50 $h^{-1}$ Mpc of the cluster center and with
$m^*_r - 1 + k_s \le m_r \le m^*_r + 2 + k_e$.  The
$k$-corrections are applied to individual galaxies at a later stage,
so these limits guarantee that we select all galaxies that can fall within
$m^*_r - 1 \le m_r \le m^*_r + 2$. The number of galaxies selected in
the cluster region is N$_{clu}$.

\item We estimate the background contribution locally. We randomly
select ten $20'\times20'$ boxes (avoiding bad areas) in a
1.3$^{\circ}$-wide annulus, starting 3 h$^{-1}$ Mpc from the cluster
center.  Galaxies are selected within the same magnitude range as used
for computing N$_{clu}$. The median counts from the ten boxes are
scaled to the cluster area to generate the background estimate
(N$_{bkg}$). We adopt the interquartile range (IQR, which is the range
between the first and third quartiles) as a measure of the error in
N$_{bkg}$, which we term $Q_{\sigma,bkg}$. The background corrected
cluster counts (N$_{clu}$ - N$_{bkg}$) is called N$_{corr}$.

\item Next, a bootstrap procedure is used to statistically apply
k$-$corrections to the galaxy populations in each cluster. In each of
100 iterations, we randomly select N$_{corr}$ galaxies from those
falling in the cluster region (N$_{clu}$). An elliptical
k$-$correction is applied to 80$ \%$ of the N$_{corr}$ galaxies, while
an Sbc k$-$correction is applied to the remaining 20$ \%$. Finally, we
use these k$-$corrected magnitudes to count the number of galaxies
with $m^*_r - 1 \le m_r \le m^*_r + 2$. The final richness estimate
N$_{gals}$ is given by the median counts from the 100 iterations. The
richness error from the bootstrap procedure alone is given by
$Q_{\sigma,boot}$. The richness error is the combination of this
error and the background contribution, so that $Q_{\sigma}$
= $\sqrt{Q_{\sigma,boot}^2 + Q_{\sigma,bkg}^2}$. At this point, the richness error includes contributions from the $k$-correction and the background galaxy correction, but not the redshift uncertainty, which is incorporated in Step 5.

\item If the cluster is too nearby or too distant, either the bright
($m^*_r - 1 + ks$) or faint ($m^*_r + 2 + ke$) magnitude limit,
respectively, will exceed one of the survey limits ($15.0 \le m_r \le
19.5$). We then apply the appropriate incompleteness correction
to the richness estimate:

\begin{equation}
\gamma_1 = {\int_{m_r^*-1}^{m_r^*+2} \Phi(m)dm \over
\int_{15}^{m_r^*+2} \Phi(m)dm} 
\end{equation}
\begin{equation}
\gamma_2 = {\int_{m_r^*-1}^{m_r^*+2} \Phi(m)dm \over
\int_{m_r^*-1}^{20} \Phi(m)dm}
\end{equation}

\noindent We call $\gamma_1$ and
$\gamma_2$ the low and high magnitude limit correction factors.

\item The above steps are repeated using each of the $n_{successful}$
photometric redshifts. The final cluster richness is recorded as the
mean of the corresponding $n_{successful}$ richnesses. We compute the
mean of the richness errors from Step 3, as well as the dispersion
among the richnesses from the $n_{successful}$ iterations. The former
quantifies how much richness variation we expect based solely on
cosmic variance, assuming there is no error in the redshift estimates,
while the latter reflects the richness error due to scatter in the
photometric redshifts. Because these are two independent sources of
error, we add them in quadrature to derive the final richness error
$Q_{\sigma,Ngals}$.

\end{enumerate}

Total $r$-band luminosities (in solar units) and their errors are
computed similarly to the richnesses. No attempt is made to fit and
integrate luminosity functions for the individual clusters.

%For a subset of the clusters in our survey, we now also compute the
%richness $N_{200}$ within the dynamical radius $r_{200}$. Computation
%of $r_{200}$ is described in \S3.2. We compute $N_{200}$ the same way as 
%$N_{gals}$, simply replacing the fixed aperture with the one defined
%by $r_{200}$ for each cluster. The background correction for $N_{200}$
%is computed using the same annulus as for $N_{gals}$. The $N_{200}$
%richnesses are especially useful for comparison to X-ray luminosities,
%typically measured within this radius (see \S6.2).

\subsection{Completeness and Contamination}

\begin{figure}
\vskip 1.3truein
%\epsscale{0.9}
\plotone{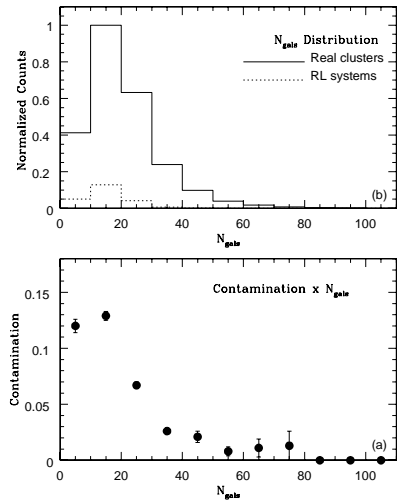}
\caption{Contamination of the entire NoSOCS catalog by false
clusters. The top panel shows the distribution of real clusters (solid
line) and false clusters (from the RL simulations, dotted line), as a
function of richness. The total contamination rate is 8.4\%,
consistent with our goal of achieving $\le10\%$ contamination when
setting the cluster detection parameters. The bottom panel shows the
contamination rate as a function of richness. For very rich clusters
($N_{gals}>50$) the contamination rate is negligible, and only rises
above 5\% for $N_{gals}<20$. }
\label{contam_allsky}
\end{figure}

The global contamination rate is estimated following
\citet{lop04}. For each plate, we use the Raleigh-Levy (RL)
distribution to generate $N_{real,i}$ x,y coordinates, where
$N_{real,i}$ is the number of galaxies in the DPOSS catalog for plate
$i$.  Each galaxy in the RL catalog is assigned a magnitude selected
randomly from the real data.  The density mapping and cluster
detection is performed on the RL catalog for each plate, and the
detected 'clusters' are assigned photometric redshifts at random from
the real clusters in that plate. To estimate the global contamination
rate for our sample, we ran this procedure on a plate-by-plate basis
for the entire NoSOCS area. The results are shown in
Figure~\ref{contam_allsky}, where the top panel shows the distribution
of real clusters (solid line) and false clusters (from the RL
simulations, dotted line), as a function of richness. The total
contamination rate is 8.4\%, consistent with our goal of achieving
$\le10\%$ contamination when setting the cluster detection
parameters. The bottom panel shows the contamination rate as a
function of richness. For very rich clusters ($N_{gals}>50$) the
contamination rate is negligible, and only rises above 5\% for
$N_{gals}<20$.

The redshift- and richness-dependent completeness functions for each
plate are provided in Table~\ref{selfn}. The first column gives the
plate number; for each plate, there are 42 entries, using six richnesses
($N_{gals} = 15, 25, 35, 50, 80, 120$) given in the second column, at
seven redshifts ($z = 0.08$ to 0.32 with $\delta z=0.04$) given in the
third column. The fourth column gives the recovery rate (in percent)
of clusters with the given richness, at the listed redshift, for that
specific plate.  To use a large cluster sample for cosmology requires
knowledge of the mass-dependent selection function (SF). Currently, there are two methods
to generate these, either by empirically calibrating a mass-observable
relation, or using large simulations to construct mock galaxy catalogs
from which clusters are selected. \citet{koe07a} use the latter to
estimate the purity and completeness of the SDSS MaxBCG cluster
catalog, but we do not have such simulations corresponding to our
data. Using X-ray observations, we are in the process of developing an
optimized richness estimator to generate an $N_{gals} -$ Mass
relation. However, that work is beyond the scope of this paper.
 
\begin{deluxetable}{crrr}
\tabletypesize{\small}
\tablecolumns{5} 
\tablewidth{0pc} 
\tablecaption{Completeness Functions}
\tablehead{
\colhead{Plate} & \colhead{Richness} & \colhead{Redshift} &  \colhead{Completeness}
}
\startdata
005 & 015 & 0.08 & 34.0 \\
005 & 015 & 0.12 & 10.0 \\
005 & 015 & 0.16 & 10.0 \\
005 & 015 & 0.20 & 2.0 \\
005 & 015 & 0.24 & 0.0 \\
005 & 015 & 0.28 & 0.0 \\
005 & 015 & 0.32 & 0.0 \\
005 & 025 & 0.08 & 78.0 \\
005 & 025 & 0.12 & 68.0 \\
005 & 025 & 0.16 & 50.0 \\
005 & 025 & 0.20 & 12.0 \\
005 & 025 & 0.24 & 8.0 \\
005 & 025 & 0.28 & 0.0 \\
005 & 025 & 0.32 & 0.0 \\
005 & 035 & 0.08 & 96.0 \\
005 & 035 & 0.12 & 82.0 \\
005 & 035 & 0.16 & 72.0 \\
005 & 035 & 0.20 & 30.0 \\
005 & 035 & 0.24 & 8.0 \\
005 & 035 & 0.28 & 8.0 \\
005 & 035 & 0.32 & 0.0 \\
005 & 055 & 0.08 & 98.0 \\
005 & 055 & 0.12 & 92.0 \\
005 & 055 & 0.16 & 92.0 \\
005 & 055 & 0.20 & 82.0 \\
005 & 055 & 0.24 & 46.0 \\
005 & 055 & 0.28 & 10.0 \\
005 & 055 & 0.32 & 6.0 \\
005 & 080 & 0.08 & 100.0 \\
005 & 080 & 0.12 & 98.0 \\
005 & 080 & 0.16 & 94.0 \\
005 & 080 & 0.20 & 92.0 \\
005 & 080 & 0.24 & 64.0 \\
005 & 080 & 0.28 & 34.0 \\
005 & 080 & 0.32 & 8.0 \\
005 & 120 & 0.08 & 100.0 \\
005 & 120 & 0.12 & 100.0 \\
005 & 120 & 0.16 & 98.0 \\
005 & 120 & 0.20 & 100.0 \\
005 & 120 & 0.24 & 90.0 \\
005 & 120 & 0.28 & 64.0 \\
005 & 120 & 0.32 & 24.0  \\
\enddata 
\label{selfn}
\end{deluxetable} 

\subsection{Other Changes}
Following Paper II, we generate ten additional AK maps for each plate,
using a set of galaxy catalogs for each plate with random photometric
zero-point offsets added to the $r$-band magnitudes, drawn from the
known photometric error distribution for DPOSS given in
\citet{gal04}. In Paper II we required that a cluster candidate be
detected in seven of the ten zero-point-error-added maps {\em in
addition} to the original map. In the final catalog, we now require
only that a candidate be detected in {\em any} seven of the eleven
maps, as there is no {\em a priori} reason to give preference to the
original map.  This typically results in $\sim5-10\%$ additional
candidates per plate.

\begin{deluxetable*}{lrccccccccc}
\tabletypesize{\small}
\tablecolumns{8} 
\tablewidth{0pc} 
\tablecaption{Statistics From Three NoSOCS Regions}
\tablehead{
\colhead{Region} & \colhead{N$_{clusters}$} & \colhead{Area ($deg^2$)} &  \colhead{$\rho$ (N deg$^{-2}$)} & \colhead{$z_{med}$} & \colhead{$z_{Q\sigma}$} & \colhead{$N_{gals,med}$} & \colhead{$N_{gals,Q\sigma}$} & \colhead{P($z_{phot}$)} & \colhead{P($N_{gals}$)} 
}
\startdata 
NGP, good & 7985 & 5681.31 & 1.405 & 0.1416 & 0.057 & 18.65 & 9.73 & -- & -- \\
NGP, poor & 3491 & 2812.54 & 1.241 & 0.1442 & 0.057 & 18.96 & 9.64 & 0.259 & 0.326 \\
SGP       & 4026 & 2917.28 & 1.380 & 0.1309 & 0.061 & 17.83 & 9.72 & 0.000 & 0.312 \\
Combined  & 15502 & 11411.13 & 1.358 & 0.1393 & 0.058 & 18.50 & 9.75 & -- & -- \\
\enddata 
\label{3regions}
\end{deluxetable*}

\begin{deluxetable*}{crrrrrrrrrrr}
\setlength{\tabcolsep}{0.04in}
\tabletypesize{\scriptsize}
\tablecolumns{12}
\tablewidth{0pc}
\tablecaption{Cluster Properties}
\tablehead{
\colhead{Name} &\colhead{RA} &\colhead{Dec} &\colhead{$N_{det}$} &\colhead{$<z_{phot}>$} &\colhead{$\sigma(z_{phot})$} &\colhead{$N_{gals}$} &\colhead{$\sigma(N_{g})$} &\colhead{$L_{opt}$} &\colhead{$\sigma(L_{opt})$} &\colhead{$<$off$>$} &\colhead{$\beta$} 
} 
\startdata
NSC J000016+103643 &   0.06709 &  10.61203 & 11 & 0.1319 & 0.0109 &  18.2 &   4.9 & 0.468 & 0.173 & 0.17 &  29.50 \\ 
NSC J000018+204800 &   0.07902 &  20.80006 & 10 & 0.0901 & 0.0060 &   8.7 &   4.2 & 0.113 & 0.102 & 0.10 & -33.80 \\ 
NSC J000020+210327 &   0.08433 &  21.05770 &  7 & 0.1674 & 0.0014 &  23.0 &   3.6 & 0.647 & 0.230 & 0.16 & -21.80 \\ 
NSC J000024+142904 &   0.10351 &  14.48461 & 11 & 0.1254 & 0.0051 &  25.5 &   5.7 & 0.574 & 0.167 & 0.10 &  50.20 \\ 
NSC J000029+215512 &   0.12189 &  21.92005 & 11 & 0.1332 & 0.0035 &  30.1 &   4.0 & 0.550 & 0.152 & 0.05 & -13.10 \\ 
NSC J000032+141432 &   0.13691 &  14.24238 & 11 & 0.0839 & 0.0150 &  12.1 &   6.1 & 0.302 & 0.152 & 0.48 & -23.50 \\ 
NSC J000038+063046 &   0.16178 &   6.51303 & 11 & 0.2276 & 0.0063 &  33.8 &   4.3 & 1.201 & 0.565 & 0.42 & \nodata \\ 
NSC J000040+065659 &   0.16740 &   6.94999 &  7 & 0.1389 & 0.0038 &  19.6 &   4.0 & 0.383 & 0.149 & 0.09 &  38.40 \\ 
NSC J000048+125623 &   0.20284 &  12.94000 & 11 & 0.1087 & 0.0044 &  15.1 &   4.1 & 0.234 & 0.109 & 0.27 &  53.10 \\ 
NSC J000051+152013 &   0.21461 &  15.33697 & 11 & 0.1254 & 0.0044 &  18.6 &   6.3 & 0.318 & 0.139 & 0.35 &  66.90 \\ 
NSC J000056+004551 &   0.23454 &   0.76427 & 11 & 0.2114 & 0.0298 &  34.5 &   3.5 & 1.186 & 0.377 & 0.27 & \nodata \\ 
NSC J000057+064615 &   0.24058 &   6.77092 &  7 & 0.2118 & 0.0022 &   0.0 &   0.0 & 0.000 & 0.000 & 0.09 & \nodata \\ 
NSC J000105+023236 &   0.27109 &   2.54333 & 11 & 0.0927 & 0.0017 &  18.5 &   4.2 & 0.420 & 0.147 & 0.13 &  16.00 \\ 
NSC J000125+181149 &   0.35463 &  18.19711 & 11 & 0.1185 & 0.0020 &  16.4 &   3.6 & 0.262 & 0.122 & 0.10 & -19.30 \\ 
\enddata
\label{allclusters}
\end{deluxetable*}

\section{Cluster Morphological Properties}

\subsection{Substructure Measures}
Four substructure measures are computed for each candidate cluster
\citep{lop06}. Only clusters at $0.069\le z_{phot} \le 0.196$ are
examined; in this redshift range, we completely sample the cluster
luminosity function spanning $m^*_r-1 \le m_r \le m^*_r+1$. We apply
the angular separation test (AST), the Fourier elongation test
(FE), the Lee statistic (Lee 2D), and the symmetry test ($\beta$)
to 10,575 clusters, within a radius of $1.5h^{-1}$ Mpc around the
recentered positions, and a significance level threshold of 5\%. The
rationale for these choices are discussed in \S5 of \citet{lop06},
while detailed descriptions of all four tests are provided by
\citet{pin96}. Very briefly, the values taken on by the four tests indicate substructure as follow:
\begin{itemize}
\item $\beta$: For a symmetric distribution
$<\beta> \approx 0$, while values of $<$$\beta$$>$ greater than 0
indicate asymmetries.
\item AST: This statistic takes on values near
unity for substructure-free systems, and less than 1.0 for clumpy
distributions.
\item FE: Values of this statistic greater than 2.5 indicate significant deviations from circularity.
\item Lee 2D: Larger values of this statistic indicate the presence of two subclumps in the galaxy distribution.
\end{itemize} 
The main data table (Table ~\ref{allclusters})
includes only the $\beta$-test results, while Table ~\ref{subtable}
provides the results of all four tests. As noted in \citet{lop06}, the
$\beta$ test is the most sensitive to substructure.
\begin{deluxetable}{lrrrr}
\tabletypesize{\small}
\tablecolumns{5} 
\tablewidth{0pc} 
\tablecaption{Substructure Measurements}
\tablehead{
\colhead{Name} & \colhead{$\beta$} & \colhead{FE} &  \colhead{LEE2D} &  \colhead{AST}
}
\startdata
NSC J062054+861617 &   -14.9 &  2.243 &  2.667 &  33.3 \\ 
NSC J074613+854032 &   -43.2 &  1.673 &  1.644 &  21.0 \\ 
NSC J054111+843927 &   -15.1 &  1.413 &  1.547 &  25.6 \\ 
NSC J065439+845907 &   -29.9 &  2.798 &  1.722 &  17.2 \\ 
NSC J054209+842633 &    18.0 &  0.989 &  1.623 &  12.8 \\ 
NSC J055822+841733 &   -38.6 &  1.560 &  2.332 &  17.8 \\ 
NSC J061210+841036 &   -35.4 &  1.629 &  1.650 &  32.7 \\ 
NSC J065407+842104 &   -32.4 &  0.485 &  1.968 &  19.5 \\ 
NSC J064833+841519 &    84.5 &  1.180 &  2.259 &  14.8 \\ 
NSC J073249+841701 &     1.2 &  0.913 &  1.676 &  22.1 \\ 
NSC J093356+845601 &   -17.5 &  1.395 &  1.296 &  15.8 \\ 
NSC J094741+844440 &     7.0 &  0.991 &  1.386 &  21.8 \\ 
NSC J094540+843709 &     2.8 &  2.423 &  2.214 &  17.6 \\ 
NSC J083925+835412 &   -15.5 &  1.842 &  2.180 &  27.8 \\ 
NSC J083042+824948 &    24.4 &  2.217 &  1.842 &  23.1 \\ 
NSC J085200+830113 &   -86.9 &  0.425 &  2.119 &  27.5 \\ 
NSC J091312+825157 &    17.8 &  1.785 &  2.086 &  14.0 \\ 
NSC J084436+861547 &    -1.9 &  1.703 &  2.389 &  17.9 \\ 
NSC J105955+853131 &    17.1 &  2.475 &  2.048 &  31.2 \\ 
NSC J130353+844618 &    -4.5 &  1.241 &  1.572 &  26.4 \\ 
NSC J104433+840151 &   111.6 &  1.027 & 13.896 &   8.6 \\ 
\enddata 
\label{subtable}
\end{deluxetable} 

\subsection{Estimating the Dynamical Radii}

We  attempt  to estimate  the  typical length  scale
characterizing the  virialized regions of the clusters  of our sample.
Both  the theory of  gravitational collapse  in an  expanding Universe
\citealt[e.g.]{gun72} and N-body simulations  suggest that the
virialized mass of a cluster is generally contained inside the surface
where  the mean  interior  density  is about  200  times the  critical
density, $ \rho_c(z)$, at the redshift of the cluster \citep{car97}:

\begin{equation}
\left<\rho_{M}\right>_{R_{200}}  = 200 \rho_c(z)  = \frac{200}{~\Omega_M}~ \overline \rho_M(z)
\end{equation}

\noindent where $\left<\rho_{M}\right>_{R} $ is the mean mass density
of the cluster within R and $\overline \rho_M(z) $ is the mean mass
density of the Universe at redshift $z$. We assume that the radial
distribution of galaxies within a cluster follows the dark matter and
neglect possible variations of the \emph{mean} mass of galaxies,
$\overline m_{gal}$, with environment. With these simplifications,
$\left<\delta \rho_M / \overline \rho_M \right> = \left<\delta
\rho_{gal} / \overline\rho_{gal}\right> \simeq \left<\delta \nu_{gal}
/ \overline \nu_{gal}\right>$, where we define the number density of
galaxies, $\nu_{gal} = \rho_{gal}/ \overline m_{gal}$. Thus, from
Eq. 4 we get a simple formula relating the number density to
$R_{200}$:

\begin{equation}
\left<\nu_{gal}\right>_{R_{200}} \simeq \frac{200}{~\Omega_M}~ \overline \nu_{gal}(z) 
\end{equation}

{\noindent  Finally,  the  spatial  mean number  density  of  galaxies
$\left<\nu_{gal} \right>$ appearing in this formula may be related to
the        observed       \emph{projected}        number       density
$\left<\Sigma_{gal}\right>$,      through      the      approximation:
$\left<\nu_{gal} \right>_R \sim  \pi R^2 \left<\Sigma_{gal}\right>_R /
({4 \pi}/{3} ) R^3 \equiv \frac{3}{4} \left<\Sigma_{gal}\right> / R$.}

\begin{figure}
\plotone{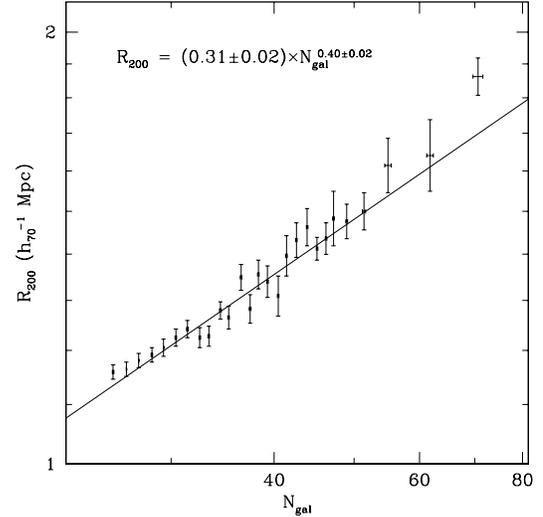} 
\caption{$R_{200}$ measurements plotted against richness for the ``best''
sample. The data were binned, subject to a minimum number of
objects/bin, and medians were taken. The errors bars give the $1\sigma$ range of values for $R_{200}$ in each bin. The best fit relation between $R_{200}$ and richness is shown.}
\label{r200fig}
\end{figure}

We calculate $R_{200}$ within the same redshift range used for
substructure measurements ($0.069<z<0.196$), but further limited to
$N_{gals}>25$ (2681 clusters). We then select only those clusters with
less than 10\% of their area within a circle of radius 1.5$h^{-1}$ Mpc
intersected by projected circles from neighboring clusters, leaving
1637 clusters.  The 10\% overlap criterion avoids structures whose
projected profiles and background regions are likely contaminated by
galaxies from a neighboring cluster. Applying our methods iteratively
could be used to relax this criterion but we have not done so as the
uncertainties on $R_{200}$ are already significant. The luminosity
function derived by \citet{bla01}, integrated to the completeness
limit of the NoSOCS catalog, $M_r = -19.8$, gives a good estimate of
the mean number density of galaxies in the Universe at redshifts
$\leq$ 0.2.  Since the NoSOCS counts are complete only in a restricted
apparent magnitude range of $15.0 \leq r \leq19.5$, for each cluster
we computed a completeness correction factor considering the absolute
magnitude limit above. These were estimated using the luminosity
function given by \citet{pao01}, which was derived from a sample of
Abell clusters detected in the DPOSS survey.
%A standard $\Lambda-$CDM cosmology were supposed, with
%$\Omega_{\Lambda} = 0.7$ , $\Omega_{M} = 0.3$ and ${\rm H_0 = 70 ~ km
%s^{-1} Mpc^{-1}}$

Following the procedure described by \citet{lop06}, for each cluster,
the background density contribution was calculated using an annular
ring about the cluster center with inner and outer radii of $R_{in} =
3 ~\rm Mpc$ and $R_{out} = 4.6~\rm Mpc$, respectively.  The cumulative
projected number density profiles appearing in Eq. 5 are then
calculated by counting galaxies in concentric annuli around the
cluster center. The ring widths are variable, defined by requiring a
constant number of galaxies per ring. These counts were then corrected
for the background contribution and for completeness. Further
corrections were applied to account for the regions within the annuli
that intersected the bad areas due to bright objects or the
densitometry spots. Furthermore, when computing $R_{200}$ for a given
cluster, areas around neighboring clusters were masked with a 1.5 Mpc
radius to avoid projection effects, resulting in large excluded areas
for low redshift clusters.  For each cluster, its galaxy number
density, was calculated by excluding galaxies located in the overlap
areas and correspondingly correcting the counting areas.  For each
cluster, the ratio $f_{overlap}$ of the number of galaxies in the
overlapping areas to the total number of galaxies within a maximum
search circle centered on the cluster center ($R_{search}$), was
estimated. Clusters with $f_{overlap} > 0.7$ did not have $R_{200}$
computed. For 17 of 1637 clusters the measurement failed because the computed values were unphysically large ($R_{200} > 4$ Mpc
$h_{70}^{-1}$, thus extending to the background area).  For 2 clusters
the background density was too high and no meaningful density profile
could be obtained.

The solution $R_{200}$ of Eq. 5 is obtained by spline interpolating
the cumulative density profile using the 5 points nearest to the
solution.  Figure~\ref{r200fig} show results for the ``best'' clusters
as a function of richness.  This subsample consists only of clusters
with high richness ($N_{gals} > 25$), chosen to reduce the effects of
background fluctuations. Furthermore, clusters whose analysis regions
were affected by bad areas or neighboring clusters over $>50$\% of
their total projected areas were discarded, as were those which
crossed plate boundaries.

Examination of the $R_{200}$ values shows that they span the same
range as those reported in \citet{han05}, but we find very large
scatter as a function of $N_{gal}$. This is seen in the error bars in
Fig.~\ref{r200fig}, which are large despite having $\sim20$ clusters
per bin. This is likely due to a combination of shallow depth, large
photometric errors and the exclusion of significant regions due to
cluster overlaps. Nevertheless, the overall relation between $R_{200}$
and $N_{gals}$ is reasonable, as shown in Fig.~\ref{r200fig}. A linear
best fit to this data with $R_{200}\propto N^{\alpha}_{gal}$ yields
$\alpha = 0.40 \pm 0.02$, well within the expectations from the
results of the analysis by \citet{lop06}. In that work it was shown
that for X-ray clusters in common with a subsample of NoSOCS clusters
without substructure, $T_X \propto N_{gal}^{\beta}$, with $\beta \sim
0.8$. Since $M(R_{200}) \propto T^{3/2}$ , where $M(R_{200})$ is the
cluster mass inside $R_{200}$, it follows that $R_{200} \propto
N_{gal}^{\beta/2}$, as we have found here.

These findings are comparable to those in the literature. The range of
$R_{200}$ spanned by our clusters is similar to those in \citet{han05}
when transformed to their cosmology with $h=1$, although we do not
extend to the lowest richness systems ($N_{gals,MaxBCG}<4$) that they
include. They also find, using a richness measured solely from the red
sequence in the MaxBCG technique, $\alpha=0.57$ in the $r'$
band. Similarly, \citet{pop07}, examining clusters detected in both
the RASS and SDSS, find $N_{200}\propto M_{200}^{0.91}$; assuming mass
scales with volume this yields $\alpha=0.37$. \citet{col05} looked at
clusters and groups in the 2dFGRS and derived $N\propto M^{0.99}$,
which gives $\alpha=0.34$. Using K-band data, \citet{lin04} find
$N_{gal}\propto M_{200}^{0.85}$, or $\alpha=0.39$. The aforementioned
surveys compare richness and mass both measured within $R_{200}$. A
direct comparison $R_{200}$ to richness measured in a fixed physical
aperture was done by \citet{yee03}, who used a radius of 0.5 Mpc
$h_{50}^{-1}$ when examining CNOC clusters; they find $R_{200}\propto
B_{gc}^{0.47}$. As noted above, we find $R_{200}\propto
N_{gals}^{0.41}$, broadly consistent with all of these results despite
the large differences in richness measurement techniques.  The
photometric data, cluster detection and especially richness
measurements are all distinct, and a full comparison would require
running analogous detection and richness codes on both datasets. Such
bidirectional tests will be fundamental to assessing systematic
effects in cluster catalogs.

\section{Global Sample Properties}

\subsection{Comparison of Three Regions}

\begin{figure}
\plotone{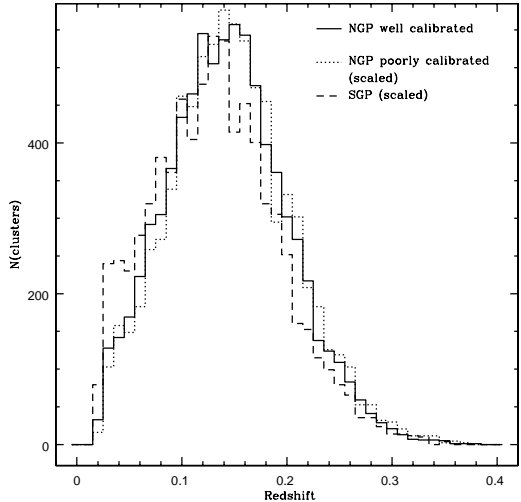}
\caption{Redshift distributions for the three independent regions of NoSOCS.\label{redcompare}. The NGP poorly calibrated (dotted line) and SGP (dashed line) distributions are scaled to the same total number of clusters as in the well-calibrated NGP region (solid line).}
\end{figure}

\begin{figure}
\plotone{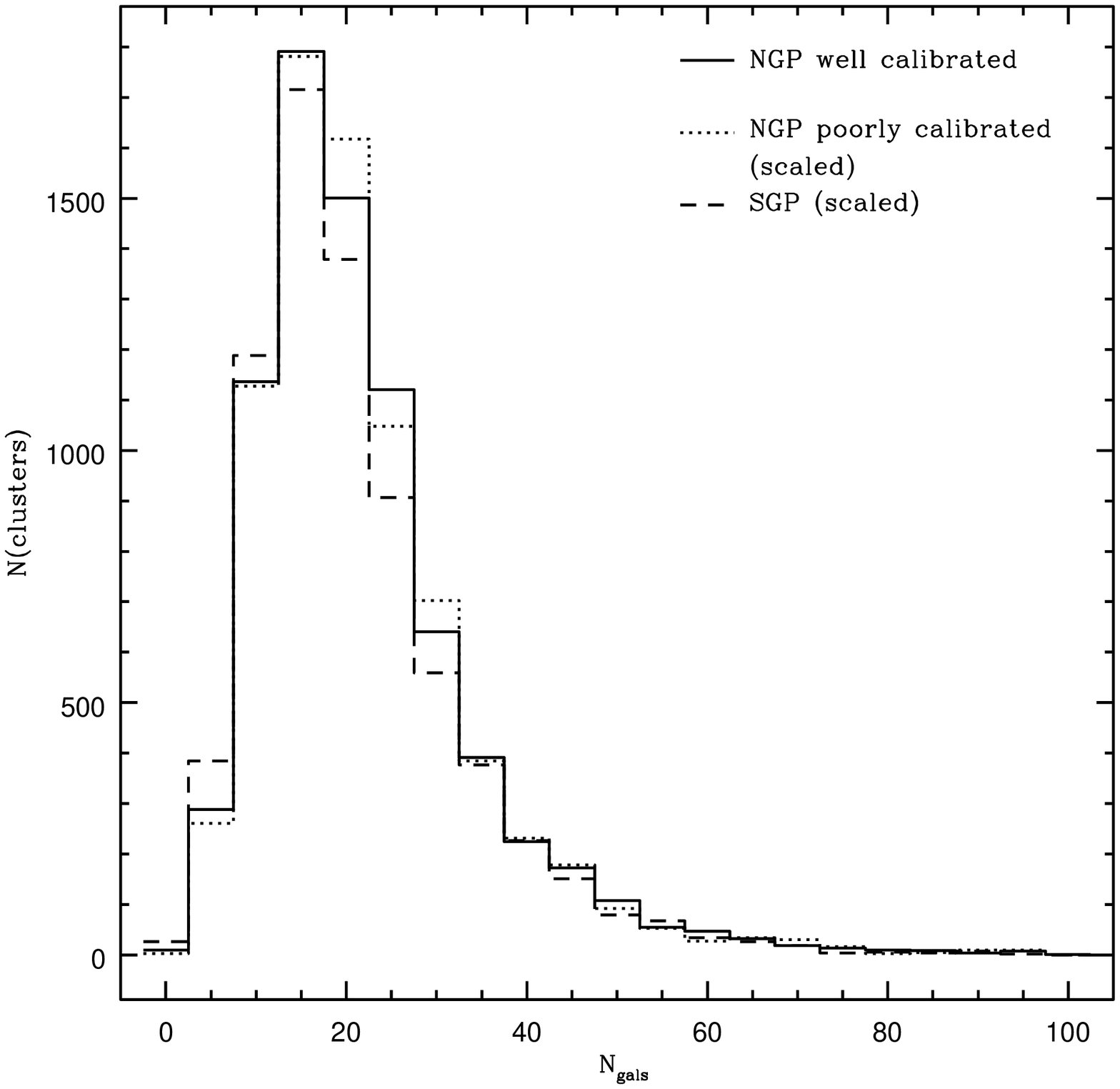}
\caption{Richness distributions for the three independent regions of NoSOCS. The NGP poorly calibrated (dotted line) and SGP (dashed line) distributions are scaled to the same total number of clusters as in the well-calibrated NGP region (solid line).
\label{richcompare}}
\end{figure}

\begin{figure}
\vskip 1.3truein
%\epsscale{0.9}
\plot3{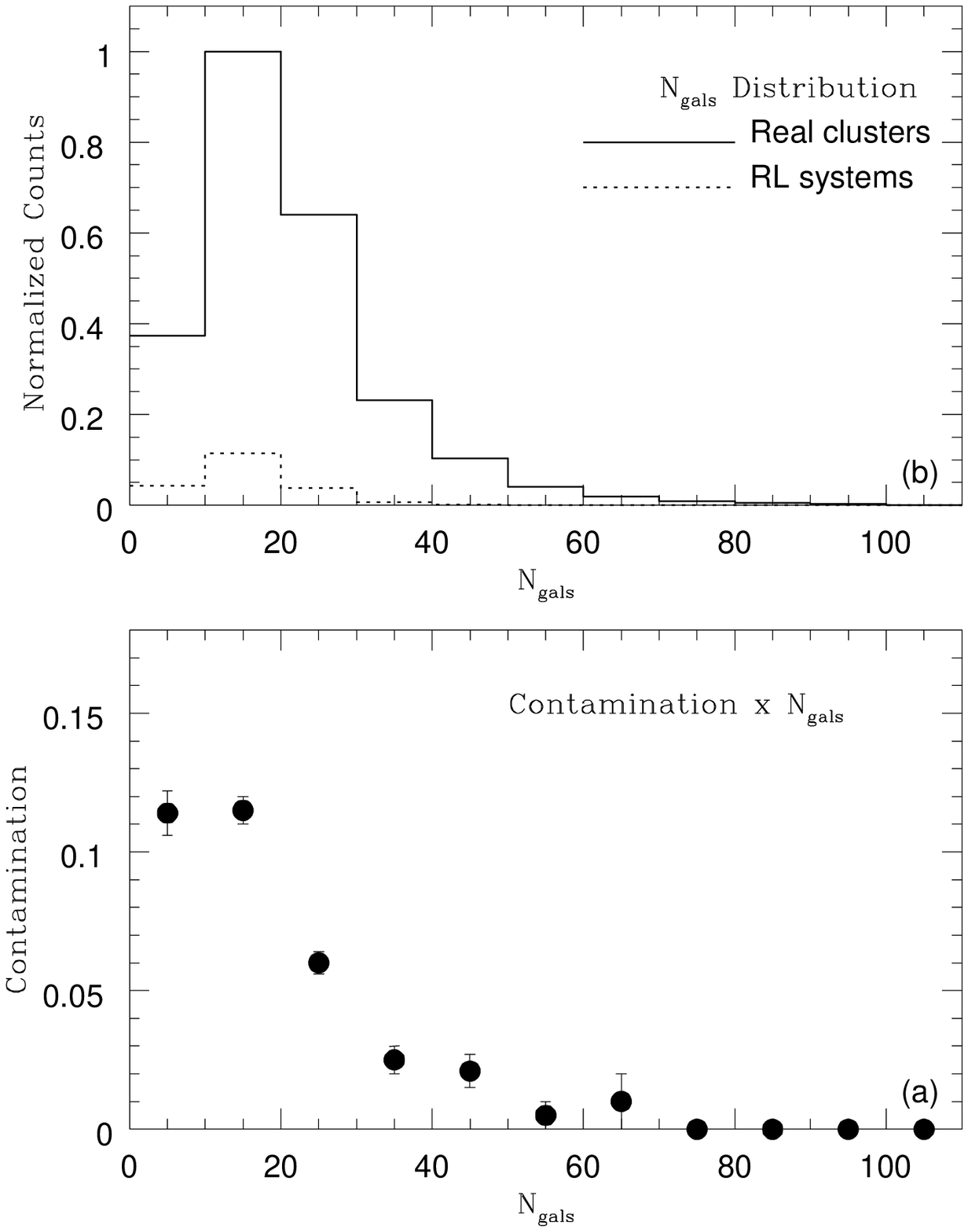}{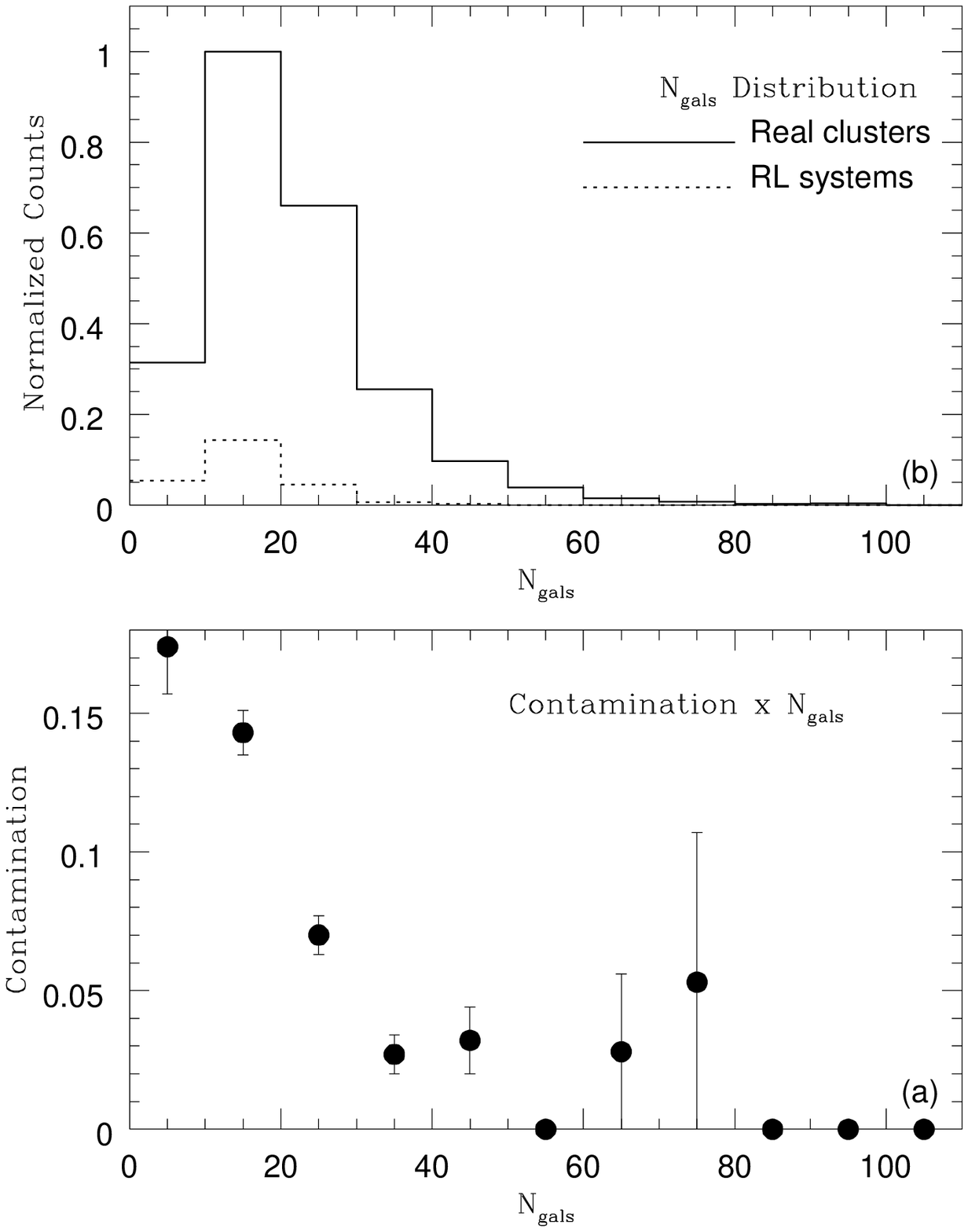}{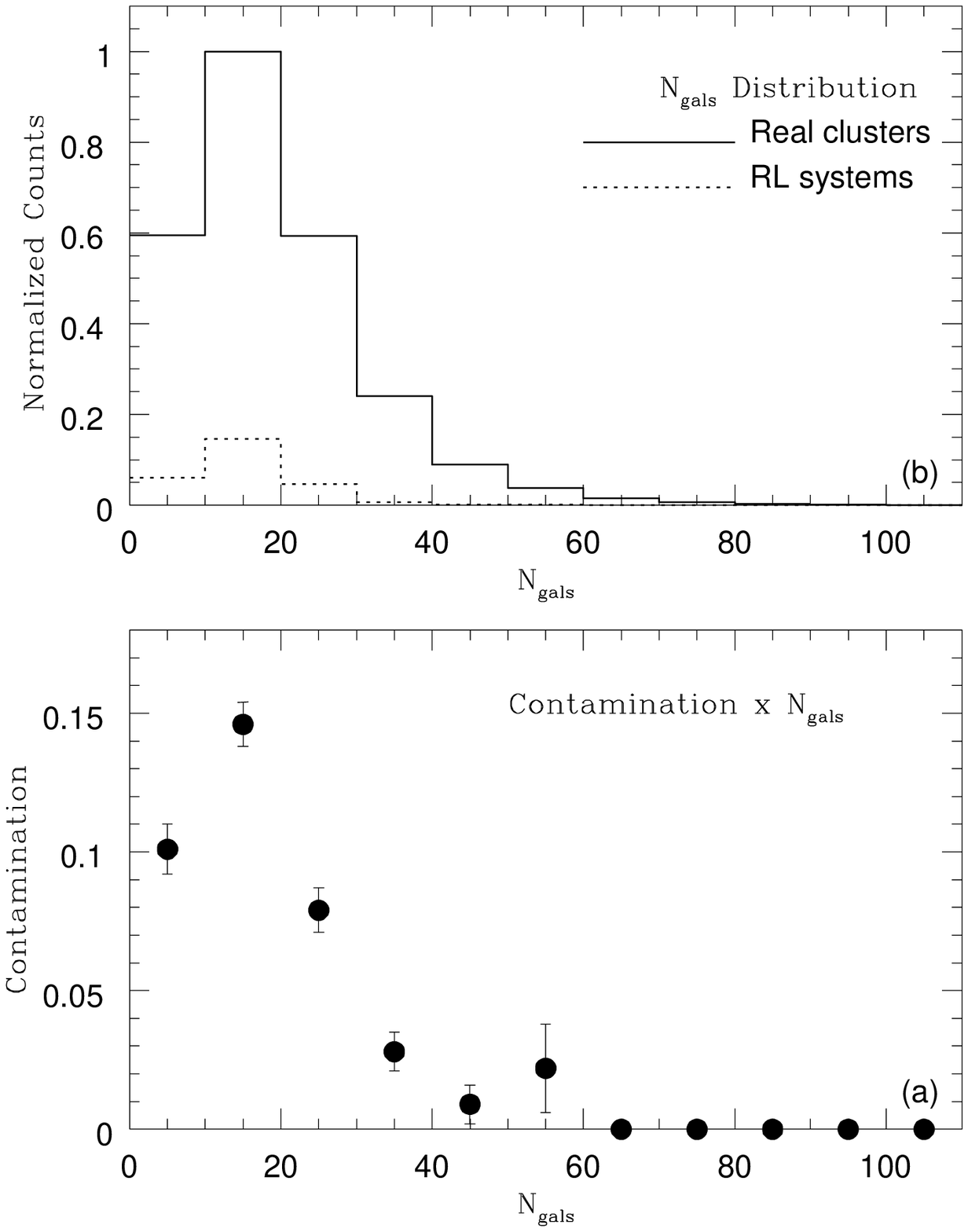}
\caption{Contamination rates in the three separate NoSOCS regions. From left to right, we show the well-calibrated NGP, poorly calibrated NGP, and SGP, while the top and bottom rows are the same as Fig.~\ref{contam_allsky}.}
\label{contam_threeareas}
\end{figure}

As discussed earlier, the cluster catalogs presented here and in Paper
II are generated for three independent regions. Although the
photometric calibration, object classification, and cluster detection
are all performed in an identical manner, one may still expect
systematic variations between these areas, especially when considering
the poorly calibrated NGP region. With the extremely large number of
cluster candidates in the three regions, we expect that the redshift
and richness distributions should be very similar. We show the results
of this comparison for redshifts in Figure~\ref{redcompare} and
richnesses in Figure~\ref{richcompare}. The histograms for all
areas are scaled to the same total number of clusters as the
well-calibrated NGP region. Table~\ref{3regions} gives the region
name, number of clusters, total area and projected density of clusters
in Columns 1 through 4, as well as the median and $Q_{\sigma}$ for
$z_{phot}$ (Columns 5 and 6) and $N_{gals}$ (Columns 7 and 8). The
$P$-values from Kolmogorov-Smirnov tests comparing the redshift and
richness distributions for both the poorly calibrated NGP region and
the SGP region to the well-calibrated NGP region are provided in
Columns 9 and 10 of Table~\ref{3regions}. We test the redshift
distributions over the range $0.07<z_{phot}<0.3$ (where the
completeness is high) and find that they are consistent for the
two NGP areas, while the SGP region is discrepant, with
an excess of clusters at $z<0.13$. Beyond this redshift, the SGP
redshift distribution agrees very well with the other regions. We also
examined the contamination rates estimated in \S2.4 separately for
each region. The results are shown in
Figure~\ref{contam_threeareas}. The three regions are evidently very
similar, although the poorly calibrated NGP area may be slightly worse, as we would expect with less accurate photometry. Whether the differences can be attributed to cosmic variance or not is unclear; structures on scales of $\sim300 h^{-1}$ Mpc, corresponding to a redshift range locally of $\Delta z\sim0.1$ are seen in 2MASS and other surveys \citep{fri03,cou04}.

\subsection{The Final Catalog}

The complete catalog of 15,502 clusters is presented in Table~\ref{allclusters}. The columns in this table are:
\begin{enumerate}
\item  Cluster Name. The name is NSC (for Northern Sky Cluster), followed by the coordinates JHHMMSS+DDMMSS. 
\item Right ascension in J2000.0 decimal degrees. For clusters where the photometric redshift estimator succeeded, this is the mean of the recentered positions. Where the photo-$z$ failed, this is the original detected position. 
\item Declination in J2000.0 decimal degrees. See notes for RA.
\item The number of times $n_{det}$ this cluster was detected in the 11 detection passes (see \S2.5).
\item The mean photometric redshift, $z_{phot}$, from the ten photo-$z$ runs.
\item The photometric redshift error, including the contribution from the scatter in the photo-z relation and the multiple photo-z runs.
\item The mean richness $N_{gals}$ from the ten richness runs.
\item The richness error, including contributions from the $k$-corrections, background variance, and redshift errors.
\item The $r$-band optical luminosity $L_{opt}$, in solar units. 
\item The luminosity error.
\item The $\beta$ substructure parameter. This was only calculated for clusters at $0.069\le z_{phot} \le 0.196$.
\item The mean offset (in Mpc) from the original detected position in the ten photo-z runs. If the photo-$z$ failed, this is left blank.
\end{enumerate}

For the subset of 2,681 clusters with $N_{gals}\ge25$ and
$0.069<z<0.196$, we provide the X-ray luminosities measured within fixed apertures of 0.5 and 1.0 $h^{-1}$ Mpc (in units of 10$^{43}$ erg s$^{-1}$) along with the associated errors and X-ray temperatures in Table~\ref{lx}. The derivations of the
X-ray quantities are discussed in \S6 below. Table ~\ref{subtable}
provides the results of all four substructure tests for 10575 clusters
at $0.069<z<0.196$.
 
\section{Comparison to The SDSS MaxBCG Catalog}

It is instructive to compare large cluster catalogs covering the same
sky area, both as a consistency check for the newer catalog and to
search for possible systematic errors. In our earlier work we compared
the first NoSOCS area to the Abell catalog. Since then, a new, deeper
cluster catalog based on Sloan Digital Sky Survey data and using the
MaxBCG algorithm has been published by \citet{koe07b}. In this section
we compare our catalog to theirs, examining recovery rates and
richness estimates. 

The Sloan Digital Sky Survey \citep{yor00} has been used to generate a
variety of cluster catalogs with different techniques, some of which
are compared in \citet{bah03}. However, that work used only a small
area of the sky with early SDSS, spanning a few hundred square
degrees. The only cluster detection technique applied to a majority of
the SDSS sky coverage is the red sequence - brightest cluster galaxy
technique called MaxBCG, described in detail in \citet{koe07a}. The
sample described in \citep{koe07b} covers $\sim7500$ deg$^2$,
containing nearly 14,000 clusters. The increased depth, higher
photometric accuracy and multiple passbands of SDSS allow for the
generation of cluster catalogs that are more complete for poor
clusters, extend to higher redshifts, and yield better photometric
redshift estimates. Furthermore, \citet{koe07a} have used cosmological
simulations to assess their completeness and false detection rate as a
function of cluster mass. Thus, their work can be used as a benchmark
for the NoSOCS sample, which covers a larger sky area.

It is important to remember that the bright flux limit of the galaxy
catalog used here makes NoSOCS an essentially flux-limited
sample. This can be seen in Fig. 3 of Paper II; the completeness of
our survey is highly richness dependent even at $z\sim0.2$. In
contrast, the SDSS photometric catalog is $\sim3$ magnitudes
deeper. The MaxBCG method relies on the E/SO ridgeline to detect
clusters, and samples such galaxies down to $0.4L_*$ out to
$z=0.4$. Thus, the MaxBCG catalog, trimmed to $z=0.3$ to reduce
photometric redshift uncertainties, provides something close to a
volume-limited sample. The completeness is near unity for all cluster
masses $>3\times10^{14} M_{\odot}$, as shown in Fig. 7 of
\citet{koe07b}. Only for poor systems is this untrue, since the limit
of $N_{ga;s,MaxBCG}>10$ imposed on the published sample will introduce
some incompleteness. The distinction between our flux-limited catalog
and the volume-limited MaxBCG catalog is evident in the mutual
recovery rates discussed below.

\begin{figure*}
%\epsscale{0.9}
\plotone{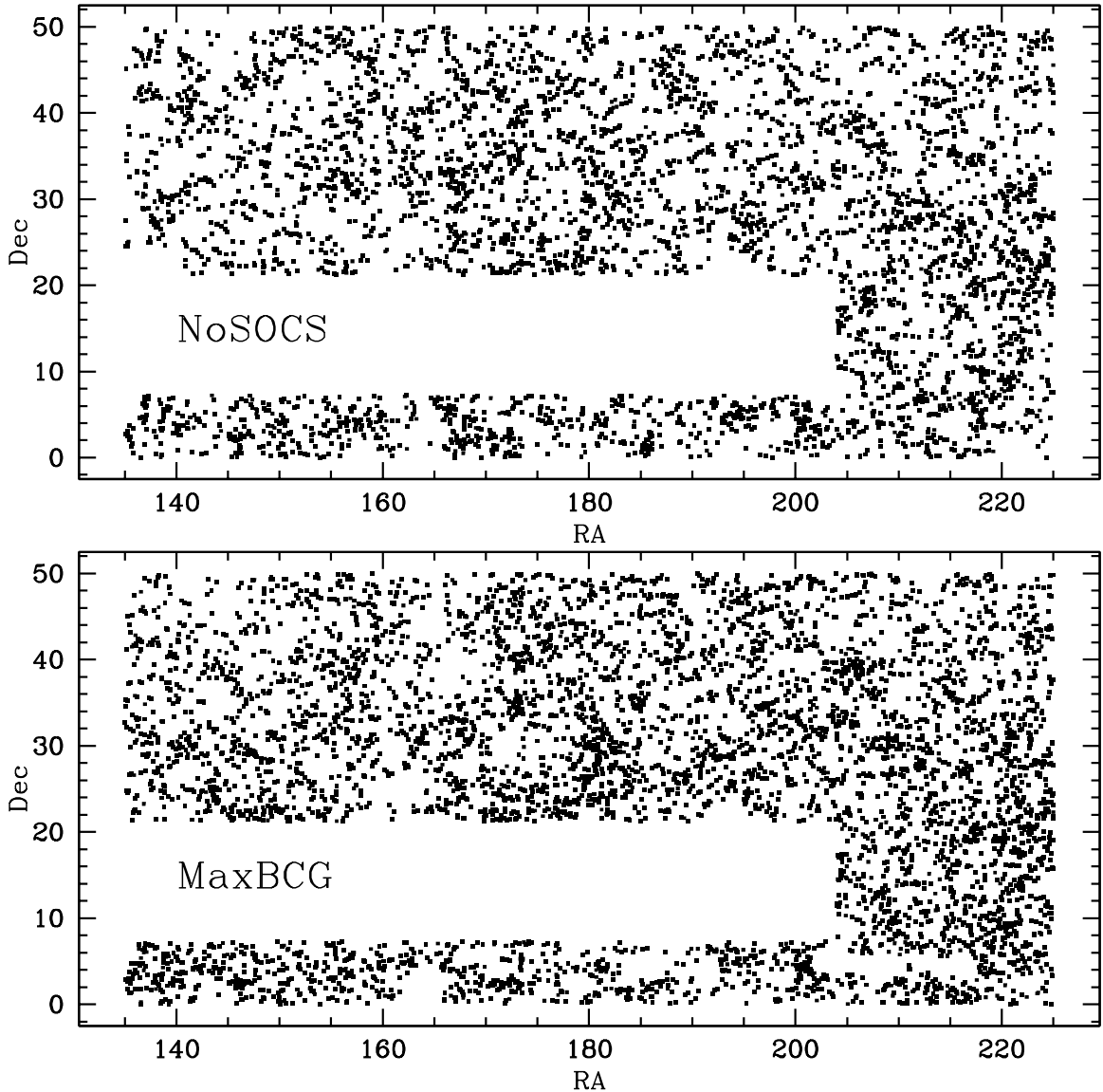}
\caption{Comparison of the projected distribution of clusters in our catalog (top) and the SDSS MaxBCG catalog (bottom). The excluded rectangular area corresponds to a stripe missing from the MaxBCG catalog. }
\label{nscvsbcg}
\end{figure*}

First, we checked the SDSS sample for clusters falling into any of our
bad areas. Only 141 of their 13,823 clusters (1.02\%) are eliminated
in this way. This suggests that the sizes of our exclusion regions are
reasonable, especially since the long exposures on the photographic
plates yield larger saturated regions around bright stars than the
shorter SDSS exposures. We restrict our comparison to the NGP region
bounded by $135^{\circ}<RA<225^{\circ}, 0^{\circ}<Dec<50^{\circ}$
since the SDSS data covers only small strips in the SGP and does not
extend to the northernmost declinations. A rectangular region is also
trimmed from both catalogs to account for a missing stripe in the SDSS
area. These cuts result in an overlap region of 3100 deg$^2$
containing 5,595 maxBCG and 4,275 NoSOCS clusters. Further restricting
our catalog to the same redshift range as MaxBCG ($0.1\le z\le0.3$)
leaves only 3,299 NoSOCS clusters. However, applying these cuts based
on our noisy photometric redshift estimator will introduce complex
effects in the comparisons, as noted by \citet{bah03}, so we do not
apply this cut to our catalog. Figure ~\ref{nscvsbcg} shows the region
of sky used, with NoSOCS clusters in the top panel and maxBCG clusters
in the bottom panel. While a generally good correspondence is seen,
there are clearly many clusters not in common to the two catalogs. The
overall large scale structure, including filaments spanning tens of
degrees, is well reproduced by both surveys.

\subsection{Does MaxBCG Find NoSOCS Clusters?}

\begin{figure}
%\epsscale{0.9}
\plotone{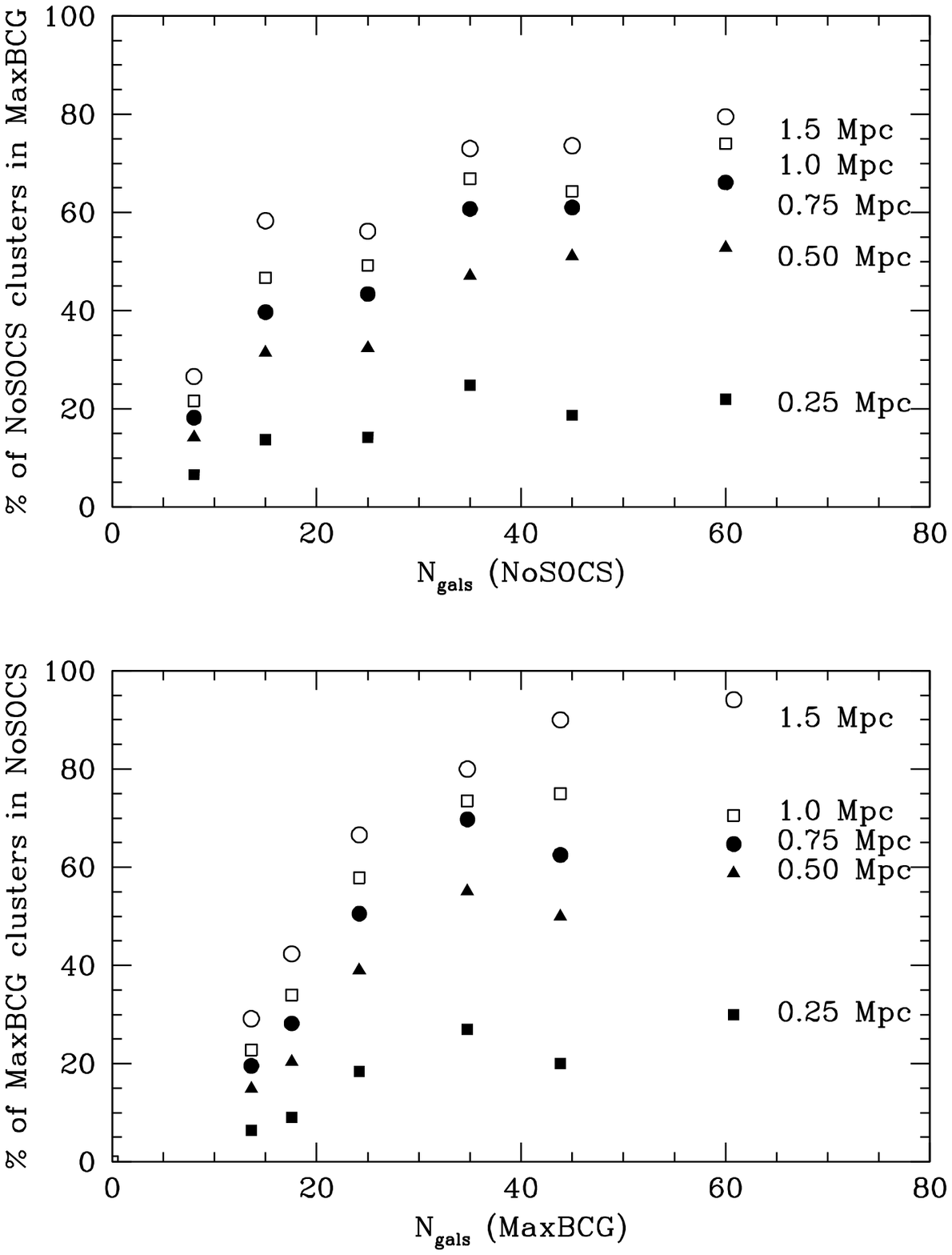}
\caption{{\bf Top:} Recovery rate of NoSOCS clusters in the MaxBCG catalog of \citet{koe07b} as a function of matching radius. {\bf Bottom:} The reverse comparison, showing the recovery of MaxBCG clusters in the NoSOCS catalog.}
\label{recoveryrates_bcg}
\end{figure}

We first examine the recovery rate of our clusters in the MaxBCG
catalog. We simply search for the closest projected match to each
NoSOCS cluster among the SDSS clusters. The angular separation between
matched clusters is converted to a physical distance in kiloparsecs
using the NoSOCS photometric redshift. Of the 4,275 NoSOCS clusters,
only 49.3\% are matched to a MaxBCG counterpart within 1 Mpc. The top
panels of Figure~\ref{recoveryrates_bcg} show the recovery rate of
NoSOCS clusters by MaxBCG as a function of matching radius and NoSOCS
richness. For poor clusters ($N_{gals,NoSOCS}<30$) the recovery rate
is low, even using large matching radii. This suggests that (a) MaxBCG
may fair poorly at detecting poor systems which have weak or no red
sequence and no BCG, and/or (b) the contamination rate in our catalog
is high for poor clusters. On one hand, the MaxBCG catalog
demonstrates a completeness of $>80\%$ for $N_{gals,MaxBCG}>10$
\citep{koe07a}, based on both Monte Carlo simulations where Abell-type
clusters are inserted into the data \citep{koe07b} as well as cluster
detection run on large mock catalogs \citep{roz07}. It is also nearly
volume-limited and should therefore contain all such structures at
$z<0.3$; however, the {\it a posteriori} limit of $N_{gals,MaxBCG}\ge
10$ imposed on the published catalog is likely to have eliminated many
poor systems that we detect. On the other hand, based on
Figure~\ref{contam_allsky}, only $\sim15\%$ of poor NoSOCS clusters
are expected to be false detections. The low recovery rate of poor
NoSOCS clusters by MaxBCG calls into question either one or both of
these results, and requires further detailed study, especially using
spectroscopic redshifts to determine the reality of these systems. We
suspect that many of the poor systems we detect but not in the
published MaxBCG catalog may have MaxBCG richnesses below their
publication threshold.

\begin{figure}
%\epsscale{0.9}
\plotone{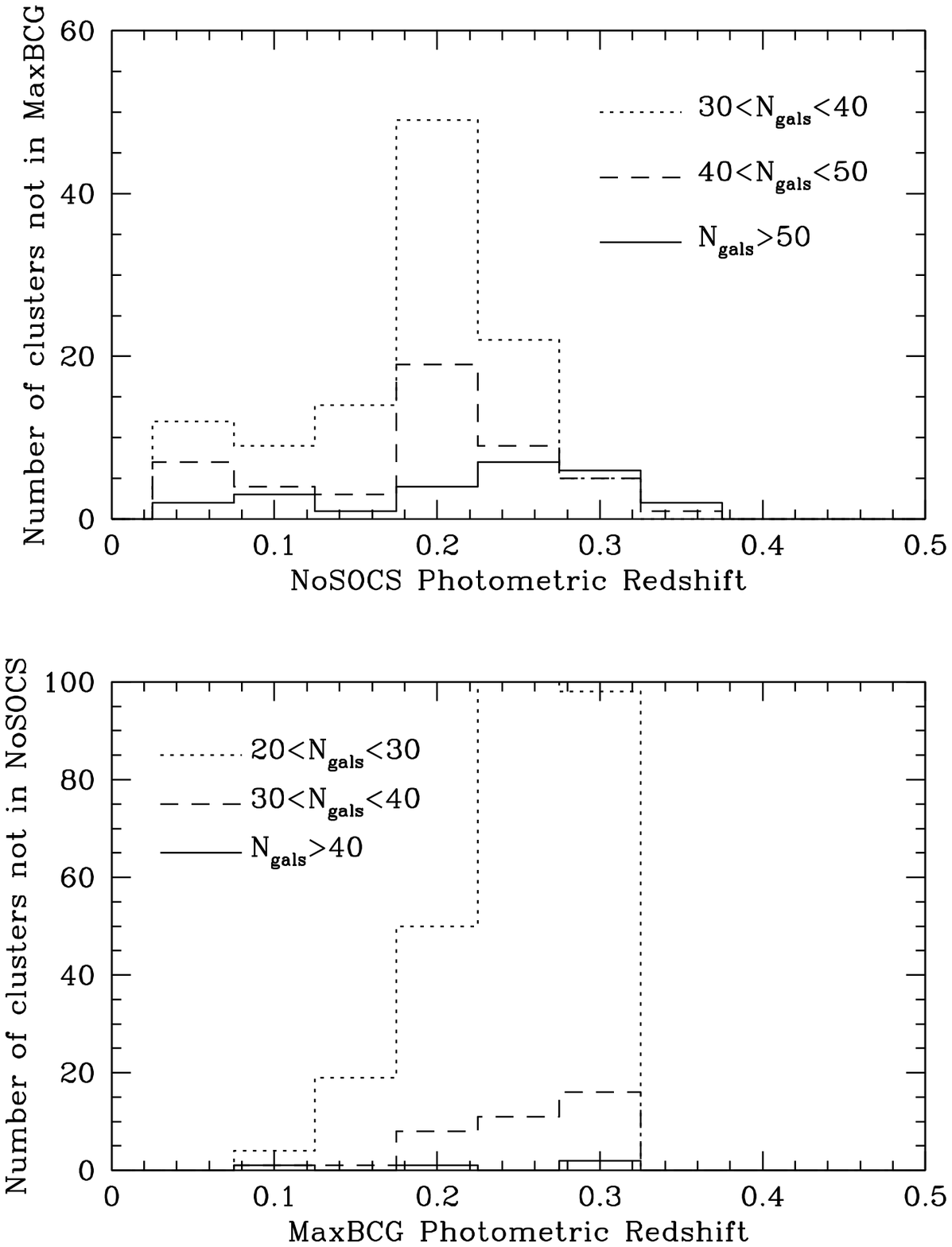}
\caption{{\bf Top:} The redshift distribution of NoSOCS clusters {\it not} found by MaxBCG. The different line types denote distinct intervals of NoSOCS richness. {\bf Bottom:} The redshift distribution of MaxBCG clusters {\it not} identified in NoSOCS. The different line types denote distinct intervals of MaxBCG richness. }
\label{notrecovered_bcg}
\end{figure}

More intriguing is the low $\sim75\%$ recovery rate for very rich
NoSOCS clusters. Our estimated completeness is $\sim95\%$ for
$N_{gals}>50$ out to $z=0.25$ \citep{gal03}, while the contamination
rate is negligible for rich clusters. Similarly, \citet{koe07a} claim
nearly 100\% completeness for similar systems. Examination of the
unrecovered systems shows that they are typically at $0.2\le
z_{phot}\le0.3$, as shown in the top panel of
Figure~\ref{notrecovered_bcg}. This suggests that a combination of the
strict {\it a posteriori} redshift limits imposed on the MaxBCG
catalog, along with the significant scatter in the NoSOCS photometric
redshifts is responsible for a significant portion of the observed
incompleteness. It is unlikely that richness errors cause this
incompleteness, since the two surveys' richnesses are well
correlated. Nevertheless, it will be important to carefully examine
the rich systems found by only one of the techniques to understand
potential biases. These comparison difficulties also show that
applying {\it a posteriori} limits (in richness, redshift, or some
other property) to publicly available catalogs makes them troublesome
(if not impossible) to use in such comparative studies.

\subsection{Does NoSOCS Find MaxBCG Clusters?}
 
Next, we reverse the sense of the comparison, examining the
completeness of our catalog relative to that of \citet{koe07b}. Here,
we use the MaxBCG catalog as the fiducial source, searching for the
nearest (in projection) NoSOCS cluster. Angular separations are
converted to physical distances using the MaxBCG photometric
redshifts. Because the MaxBCG catalog should be essentially 100\%
complete in the redshift range probed by NoSOCS, it provides a
potential basis for testing our own completeness (but see the caveats
above). The results are shown in the top panel of
Figure~\ref{recoveryrates_bcg}, with the recovery rate of MaxBCG
clusters by NoSOCS as a function of matching radius and MaxBCG
richness. It is immediately apparent that NoSOCS does extremely well
at discovering rich clusters, finding 80-100\% of the richest MaxBCG
clusters. Measured this way, NoSOCS is more complete for the richest
clusters than MaxBCG, although this may be due to clusters falling
below the $z>0.1$ limit imposed on the MaxBCG catalog. However, for
poor clusters and groups, the recovery rate is low. Nearly half of the
MaxBCG clusters have $N_{gals,MaxBCG}\le15$, falling into the lowest
richness bin in this plot. Clearly, neither our algorithm nor that of
MaxBCG is anywhere near complete for group-mass systems. The bottom
panel of Figure~\ref{notrecovered_bcg} shows that the NoSOCS
completeness drops with redshift to $z\sim0.3$, as expected for our
flux-limited survey. Nevertheless, the completeness is very high even
for moderately poor systems to $z\sim0.2$, consistent with the
estimates shown in Fig. 5 and 6 of \citet{gal03}.

\subsection{Comparison of Cluster Properties}

\subsubsection{Photometric Redshifts}
\begin{figure}
%\epsscale{0.9}
\plottwo{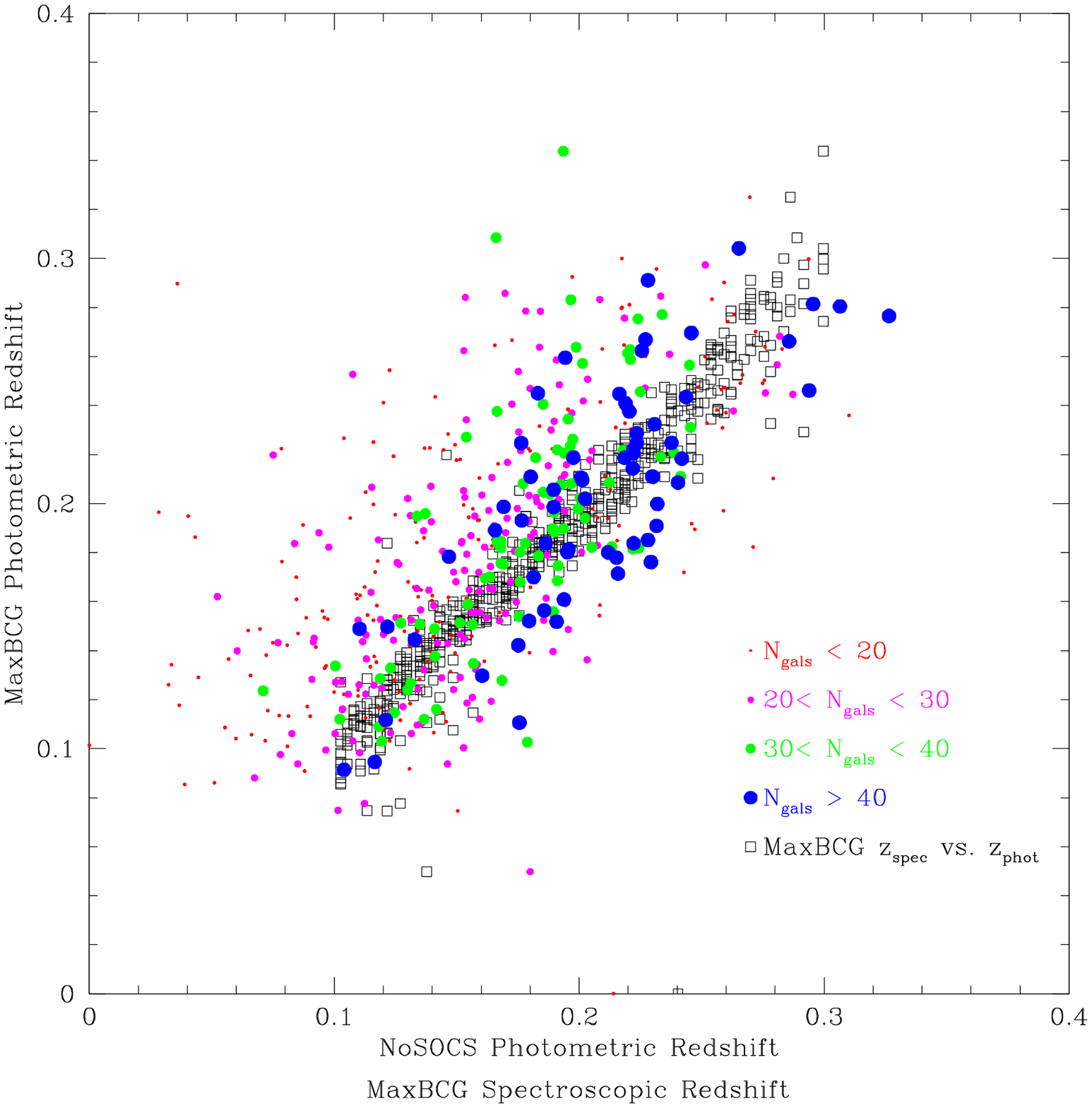}{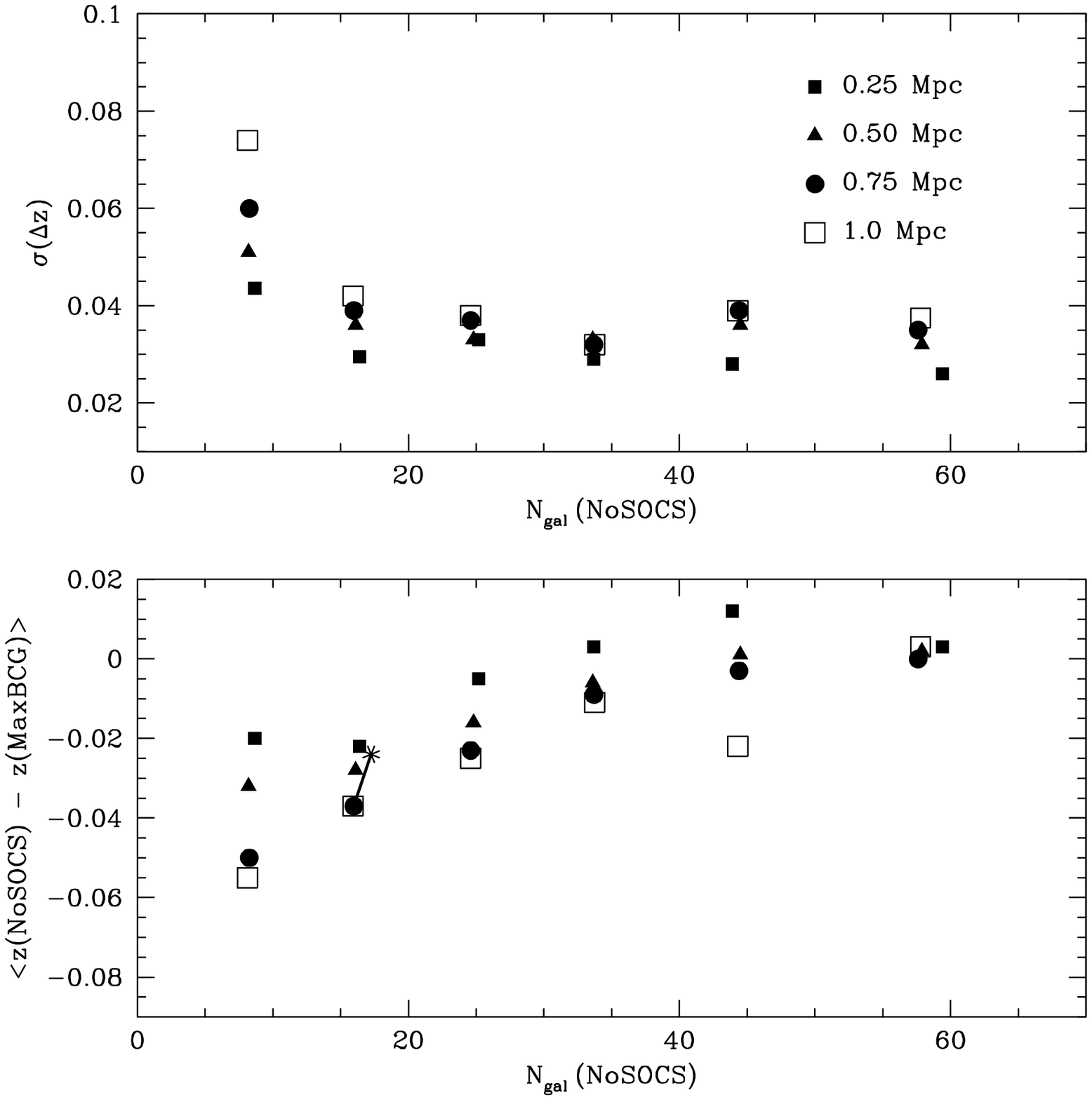}
\caption{{\bf Left:} Comparison of NoSOCS and MaxBCG redshifts for
clusters matched within 0.75Mpc. The solid circles show the NoSOCS
photometric redshifts compared to those from MaxBCG. The point size
increases with increasing richness. Open squares show the MaxBCG
spectroscopic redshifts vs. their photometric redshifts. {\bf Right:}
The dispersion (top) and median offset (bottom) between NoSOCS and
MaxBCG photometric redshifts, as a function of richness and matching
radius.}
\label{zcompare_bcg}
\end{figure}

Beyond examining the completeness of these cluster catalogs, we use
the more accurate MaxBCG photometric redshifts to test our own
estimates. We also examine the relationship between the NoSOCS and
MaxBCG richnesses. The left panel of
Figure~\ref{zcompare_bcg} shows the comparison of photometric redshift
estimators, as a function of NoSOCS richness, for NoSOCS clusters with
a MaxBCG counterpart within 0.75 Mpc. The poorest clusters are shown
as the smallest dots. Open squares show the MaxBCG photometric
redshifts on the ordinate, and their spectroscopic redshifts on the
abscissa. For poor clusters (small dots), the scatter between the two
estimators is high, and the NoSOCS $z_{phot}$ appears to underestimate
the true redshift. For clusters with $N_{gals,NoSOCS}>20$, the scatter
is dramatically reduced and there is only a small offset, which
disappears for $N_{gals,NoSOCS}>30$ . These are quantified in the
right panels of Figure~\ref{zcompare_bcg}, which shows the scatter
(top) and median offset (bottom) between the two photometric redshift
estimators, as a function of $N_{gals,NoSOCS}$ and the matching
radius. Assuming the MaxBCG measurements are more accurate, we
overestimate the redshifts of poor clusters. This may be due to the
training sample used, which consists almost exclusively of Abell
clusters, which are much richer than these poor groups. Furthermore,
because of the minimum redshift ($z>0.1$) imposed on the MaxBCG
sample, there is a bias at the low redshift end, where most of the
poor NoSOCS clusters are detected. In fact, there are only two
clusters in the sample with $N_{gals,NoSOCS}\le 10$ and $0.15\le
z_{phot,NoSOCS}\le0.25$ matched within 750kpc. If we move to the next
richness bin, $10<N_{gals,NoSOCS}\le20$, there are 115 clusters with
$0.15<z_{phot,NoSOCS}<0.25$, of which 80\% are at $z<0.19$. This
effect is shown by the line and asterisk in the bottom right panel of
Figure~\ref{zcompare_bcg}, where the median and scatter of the
$z_{phot}$ differences is computed only for those clusters with
$10<N_{gals,NoSOCS}\le20$ and $0.15<z_{phot,NoSOCS}<0.25$, compared to
the solid circle if no redshift cut is used. Applying this limited
redshift range reduced the median offset by $\sim50\%$.

\subsubsection{Richness}

The number of galaxies in a cluster may be directly related to the
underlying dark halo mass. If this is true, purely photometric cluster
surveys are adequate to construct the cluster mass function, and in
concert with photometric redshifts, measure its evolution.

\begin{figure}
%\epsscale{0.9}
\plotone{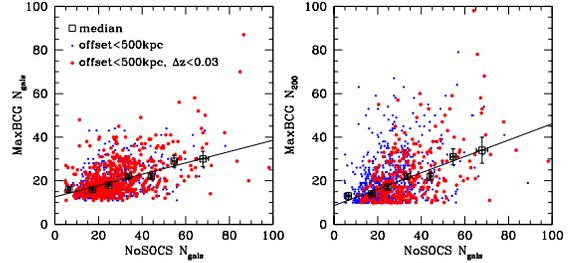}
\caption{Comparison of NoSOCS and MaxBCG richnesses for clusters with
$0.1<z_{phot}<0.3$ matched within 0.5Mpc. Small blue points show all
of these clusters, while larger red points require that the
photometric redshifts agree within $\Delta z<0.03$. Large square show
the medians in bins of width $\Delta N_{gals}=10$. The left panel
compares richnesses measured in the same fixed aperture of 500
kpc. The right panel compares our fixed-aperture richness with the
MaxBCG richness within their $R_{200}$. The best fit relations are
shown as the solid lines.}
\label{richcompare_bcg}
\end{figure}

To test the reliability of such richness estimates, we compare our
richnesses to those from the MaxBCG catalog. Both surveys compute
richnesses in fixed physical apertures as well as within
$r_{200}$. The results are show in Figure~\ref{richcompare_bcg}, where
the left panel shows our $N_{gals}$ (within a 500kpc radius aperture)
versus MaxBCG richnesses in the same aperture, and the right panel
compares our fixed-aperture richness with MaxBCG's $R_{200}$
richness. We only use clusters whose centroids agree to within 500
$h^{-1}$ kpc between the two surveys, and with $0.1<z_{phot}<0.3$,
resulting in a sample of 1,072 clusters.  The small blue points show
all matched clusters, while red points are those matches where the
NoSOCS and MaxBCG photometric redshifts differ by less than 0.03. The
large open squares show the medians in bins of width $\Delta
N_{gals}=10$, along with the rms scatter. Although there is a
moderately large dispersion between our richnesses and those from
MaxBCG, they are well correlated. The solid lines in
Figure~\ref{richcompare_bcg} show the best-fit relations:

\begin{equation}
{N_{gals,MaxBCG} = 12.55 + 0.260\times N_{gals} }
\end{equation}
\noindent and
\begin{equation}
{N_{R200,MaxBCG} = 8.60 + 0.375\times N_{gals} }
\end{equation}

As expected, the richnesses computed within the same fixed apertures
(0.5 Mpc) are much better correlated. However, these relations should
not be used to convert between richnesses from the two surveys. The
MaxBCG catalog is censored at low richness (where most of the clusters
are found) and the scatter in the relation is very high. Much of the
scatter is likely due to the different definitions of richness, where
we count all galaxies while MaxBCG effectively counts only galaxies
along the E/S0 ridgeline. Due to the large photometric errors in
DPOSS, we cannot replicate a richness using only the red-sequence
galaxies. Additional scatter is introduced by the different
photometric redshifts changing the angular sizes of the apertures
between the two surveys.

\section{X-ray Measurements}
\subsection{Comparison to NORAS}

It is instructive to compare the results of optical and X-ray cluster
surveys, for the purposes of examining completeness and testing
properties ($N_{gals},L_X,L_{opt}$) that might be useful as mass
proxies. The largest existing X-ray survey in the Northern hemisphere
is NORAS \citep{boh00}, with 378 clusters at $\delta>0^{\circ}$. NORAS
is useful not only as the largest, homogeneous catalog of
low-to-moderate redshift clusters, but also because spectroscopic
redshifts have been obtained for the entire sample. To match this
sample to NoSOCS, we first remove 55 NORAS clusters in our bad areas,
and an additional 97 at low galactic latitude. This leaves a sample of
226 NORAS clusters for comparison, of which 175 are at $0.05<z<0.3$,
where we expect NoSOCS to be very complete.

\begin{figure}
%\epsscale{0.9}
\plotone{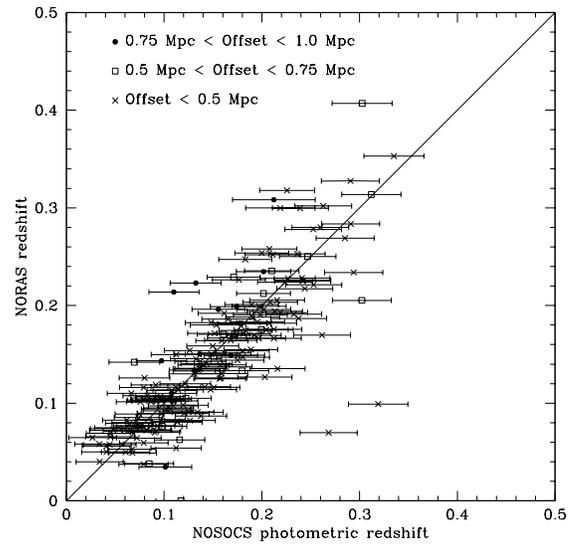}
\caption{NoSOCS photometric redshifts vs. NORAS
spectroscopic redshifts. Points are color coded by the closeness of
the positional match.}
\label{nosocsnoras}
\end{figure}

The top panels of Figure~\ref{recoveryrates_noras} show the recovery rate of
X-ray selected clusters by our survey as a function of NORAS
spectroscopic redshift (left) and X-ray luminosity (right, for
clusters with $0.05<z_{NORAS}<0.3$). At $0.05<z<0.3$ we recover
80-95\% of the NORAS clusters, depending on the matching radius, with
a distinct drop at $z>0.25$, as expected from our completeness
functions. At high $L_X$, we recover 100\% of the NORAS clusters, but
they are few in number. At moderate $L_X$ ($\sim10^{45}$ erg
s$^{-1}$), the recovery rate is quite stable near 80\%, mostly due to
clusters missed at higher redshifts. The bottom panels show the
reverse comparison, the recovery rate of optical clusters, as a
function of NoSOCS photometric redshift (left) and optical richness
(right). As with other optical cluster surveys, we have nearly two
orders of magnitude more candidates than NORAS, resulting in a very
low recovery rate of NoSOCS clusters in the X-ray. The recovery rate
increases with redshift, as the fraction of poor clusters decreases in
the optical. This effect is clearly illustrated in the bottom right
panel, as the recovery rate approaches 50\% for clusters with
$N_{gals}>80$. Nevertheless, it appears that NORAS misses over half of
the richest clusters in our sample. The extensive spectroscopy for
NORAS also allows an additional check on our photometric
redshifts. The left panel of Figure~\ref{nosocsnoras} plots NoSOCS
$z_{phot}$ against NORAS $z_{spec}$, with matches within 0.5$h^{-1}$
Mpc shown in black, $0.5<{\rm offset}<0.75 h^{-1}$ Mpc in green, and
$0.75<{\rm offset}<1.0h^{-1}$ Mpc in red. For the 145 clusters matched
within 0.5$h^{-1}$ Mpc, we find $Q_{\sigma}(z_{NoSOCS,phot} -
z_{NORAS,spec})/(1+z_{NORAS,spec}) = \mathrm{\Delta}z=0.026$,
consistent with the errors estimated from the photometric relation
combined in quadrature with the background estimation errors.  The
recovery rates and typical offsets are also in good agreement with
\citet{lop06}, who compared NoSOCS clusters to the more heterogeneous
BAX database.

\begin{figure}
%\plotone{fig15.eps}
\caption{{\bf Top Left}: NoSOCS recovery rate of NORAS clusters as f($z_{NORAS}$). {\bf Top Right}: NoSOCS recovery rate of NORAS clusters as f($L_X$), for clusters with $0.05<z_{NORAS}<0.3$. {\bf Bottom Left}: NORAS recovery rate of NoSOCS clusters as f($z_{phot}$). {\bf Bottom Right}: NORAS recovery rate of NoSOCS clusters as f($N_{gals}$).  \label{recoveryrates_noras}}
\end{figure}

\subsection{X-ray Luminosities from RASS}

As seen in Figure~\ref{nosocsnoras}, X-ray measurements from NORAS are
only available for a very small fraction of optically selected
clusters.  However, one can use the locations of optically selected
cluster candidates to measure X-ray fluxes and luminosities from RASS,
allowing us to improve the optical--X-ray correlations. Even though
the significance of X-ray emission in these areas may be too low to
identify extended sources in RASS, we can derive either fluxes or
upper limits following \citet{boh00} to generate a much larger
dataset.

We first restrict the NoSOCS sample to clusters at $0.069\le
z_{phot}\le0.196$, the same redshift range used for substructure
measurement. The optical richnesses are most reliable at these
distances, where the luminosity function is completely sampled over
the magnitude range used to derive $N_{gals}$. We further restrict the
sample to clusters with $N_{gals}\ge25$, as we do for measuring
$R_{200}$, to avoid poor clusters/groups where the X-ray emission is
unlikely to be detected or may be dominated by the X-ray halo of a
single galaxy. The X-ray luminosities $L_X$ are estimated from count
rates in ROSAT PSPC images taken as part of the RASS. Images and
exposure maps in the 0.4-2.4 keV band are retrieved from the ROSAT
archive via FTP. We avoid the softest ROSAT channels (0.1-0.4 keV)
since the background is higher. To make our measurements comparable to
those of \citet{boh00}, we follow an almost identical procedure.  The
background is estimated in an annulus with an inner radius of $20'$
and width of $21.3'$, divided into twelve sectors. The median count
rate from these sectors is used as the background, after removing any
sectors containing point sources. The details of this procedure are
described in \S3.1 of \citet{boh00}.  The cluster X-ray flux is then
computed using fixed apertures of 0.5 and 1.0 $h^{-1}$ Mpc radius. We
do not perform the growth curve analysis (GCA) because the vast
majority of NoSOCS clusters have very low X-ray fluxes, making the GCA
extremely unstable.  The computed X-ray fluxes are then corrected for
flux missing from the faint outer regions using the technique
described in \S3.5 of \citet{boh00}. Finally, only clusters whose
total counts are $3\sigma$ above the background are considered
reliable. All others are reported as upper limits. The X-ray
luminosities measured within fixed apertures of 0.5 and 1.0 $h^{-1}$
Mpc (in units of 10$^{43}$ erg s$^{-1}$) along with the associated
errors and X-ray temperatures (in keV) can be found in
Table~\ref{lx}. Column 1 gives the cluster name, while Columns 2-4
give the X-ray luminosity, luminosity error, and temperature, all
derived within a 0.5 $h^{-1}$ Mpc aperture. Columns 5-7 provide the
same quantities measured using a 1.0 $h^{-1}$ Mpc
aperture. Luminosities marked with a ``:'' are upper limits.

\subsubsection {Validation with NORAS and REFLEX}

To test our methodology, we have recomputed $L_X$ for the entire NORAS
and REFLEX \citep{boh01} samples using our software. We use the
GCA-derived apertures reported in those two surveys, along with the
redshifts, missing flux corrections and plasma models taken directly
from the respective samples. Rather than transform between the
different cosmologies used in NORAS and REFLEX, we perform all
calculations with the cosmological parameters used in those
surveys. To convert the measured total count rate to an unabsorbed
X-ray flux in the full ROSAT soft energy band (0.1 - 2.4 keV), we use
the PIMMS tool available through NASA HEASARC. We assume a
Raymond-Smith (RS) spectrum \citep{ray77} to represent the hot plasma
present in the intracluster medium, with a metallicity of 0.2 of the
solar value and the interstellar hydrogen column density along the
line-of-sight taken from \citet{kal05} and \citet{baj05}. The plasma
temperature is estimated in two different ways. First, we use a fixed
temperature of 5 keV, which is typical for clusters \citep{mar98}, and
term the resulting luminosity $L_{X5}$. Second, we use an iterative
procedure relying on the $L_X - T_X$ relation from \citet{mar98}. We
start by calculating $L_{X5}$ and finding the corresponding
temperature, assuming an RS spectrum. This new temperature is used to
recalculate the luminosity based on an RS spectrum, and the procedure
is iterated until convergence is reached, when the change in
temperature is $\Delta T_X<1keV$, comparable to the scatter in the
$L_X - T_X$ relation. The procedure typically converges in two or
three iterations. For both luminosity measures, we apply a
$K$-correction \citep{boh00} to derive the X-ray luminosity in the
rest frame 0.1 - 2.4 keV band.

\begin{figure}
\plotone{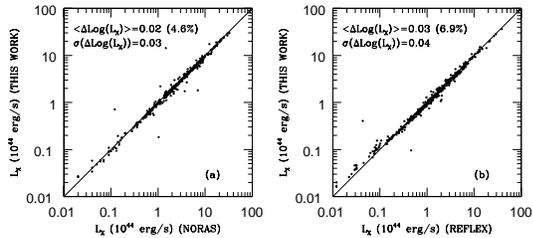}
\caption{Comparison of X-ray luminosities computed using our methodology vs. NORAS (left) and REFLEX (right) .  
\label{usvboh}}
\end{figure}

The only significant methodological differences between our technique
and the previously published works are (i) we use the 0.4-2.4 keV
images provided by the ROSAT archive, while they worked directly from
the event files in the 0.5-2 keV range, (ii) we use a metallicity of
$0.2Z_{\odot}$ instead of $0.3Z_{\odot}$, and (iii) they derive an
independent count rate to flux conversion while we rely on
PIMMS. Nevertheless, our results are in excellent agreement with both
surveys.  The comparisons to NORAS and REFLEX are shown in the top and
bottom panels of Fig.~\ref{usvboh}, respectively. We find very small
offsets of $\sim5\%$ in $L_X$ between our measurements and the
literature values, likely due to differences in the count rate to flux
conversion. The scatter is small, $\sigma(\delta Log L_x)\sim0.03$
over a very broad range of $L_X$, demonstrating that we are able to
correctly recover X-ray luminosities with our technique.

\begin{figure}
\plotone{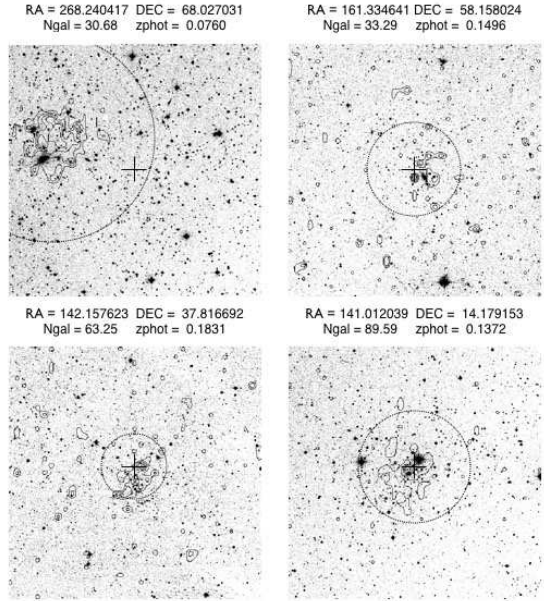}
\caption{Contours of X-ray emission from the RASS overlaid on DPOSS
$F$-band images of four clusters in our catalog. The
images are 0.5$^{\circ}$ on a side, centered on the NoSOCS optical cluster centers. There are 6 X-ray
contours evenly spaced between the background level and $2\sigma$ over
the background. A circle of radius $0.5\times R_{Abell}$ is plotted at
the X-ray flux centroid, while the optical center is marked with a cross.}
\label{optxray}
\end{figure}

Four examples of the optical and X-ray properties are shown in
Figure~\ref{optxray}, where we overlay contours of X-ray emission from
the RASS on DPOSS $F$-band images of four clusters in our catalog. The
images are 0.5$^{\circ}$ on a side, centered on the NoSOCS
optically-selected cluster centers. The clusters range in richness
from $N_{gal}=30-90$, and redshifts of $z=0.07-0.183$. There are 6
X-ray contours evenly spaced between the background level and
$2\sigma$ over the background. A circle of radius $0.5\times
R_{Abell}$ is plotted, centered on the X-ray flux centroid. Although
evident in the optical images, the X-ray fluxes are clearly not very
high, and even moderately rich clusters near the median redshift of
our catalog (such as the one at top right in the figure) do not stand
out strongly. The X-ray contours are usually well matched to the
optical center, except for the top left cluster. Visual inspection of
the galaxy distribution in the latter field shows that the NoSOCS
cluster center is in between two apparent overdensities which have
been blended in our catalog, and only one of which is X-ray
detected. This suggests that searching for clusters with highly
discrepant optical and X-ray positions and/or fluxes can be used to
find such projections.

\subsubsection{Optical vs. X-ray Properties}
\begin{figure}
\plotone{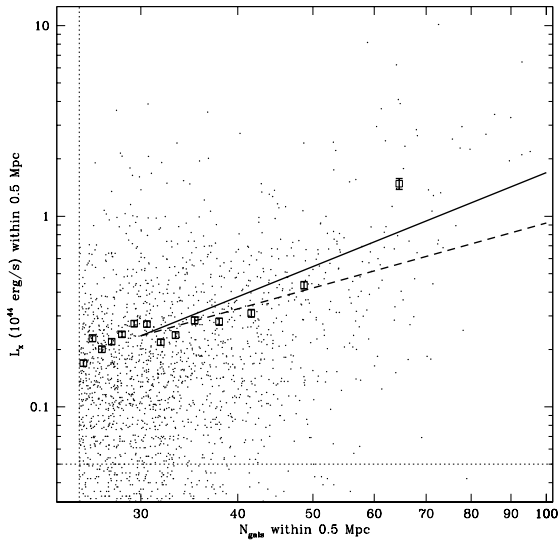}
\caption{Comparison of optical richness $N_{gal}$ and X-ray luminosity $L_X$ within 0.5 $h^{-1}$ Mpc for 1649 clusters with $0.07 < z < 0.19$, $N_{gal} >25$ and $L_X>5\times10^{42}$ ergs s$^{-1}$. Individual clusters are plotted as dots, while the binned results (with each bin containing 200 clusters)
along with their rms errors shown as the large squares with error bars. The solid line shows the relation $L_{X,750kpc}=e^{3.4}(\frac{N}{40})^{1.61}$ found by \citet{ryk08}, directly replacing their $N_{200}$ with our $N_{gals}$. The dashed line shows the same relation, but now transforming our $N_{gals}$ to MaxBCG  $N_{200}$ using Eqn. 7. Dotted lines show the $N_{gal}>25$ limit for attempting to measure $L_X$ and the $L_X>5\times10^{42}$ ergs s$^{-1}$ limit imposed on the sample when computing binned medians.}
\label{ngallx}
\end{figure}

The comparison of optical richness and our estimate of $L_X$ from the
iterative procedure described above is shown in Figure~\ref{ngallx},
using 1649 clusters with $0.07 < z < 0.19$, $N_{gal} > 25$ and
successfully measured X-ray luminosities with $L_X>5\times10^{42}$
ergs s$^{-1}$. Clusters where the X-ray luminosity is only an upper
limit and those where the background estimation failed are not
included. Individual clusters are plotted as dots, while the binned
results (with each bin containing 200 clusters) along with their
$1\sigma$ scatter are shown as the large points with error bars.

While the scatter is large, the binned relationship agrees with that
found by \citet{lop06} using higher quality X-ray data, $L_X\propto
N_{gal}^{1.616}$. We also show, as the solid line, the relationship
between $L_X$ within 750 kpc and $N_{200}$ from the X-ray stacking
analysis performed by \citet{ryk08} using the MaxBCG cluster catalog
(their equation 5), but simply replacing $N_{200}$ with our
$N_{gals}$. The power-law slopes of the \citet{lop06} and
\citet{ryk08} relations are nearly identical, despite the completely
different richness measures. As seen in Figure~\ref{ngallx}, a similar
relationship holds for our clusters, despite ROSAT's limited spatial
resolution and count rate as well as the limited photometric accuracy
and depth of DPOSS, demonstrating that a reliable cluster sample can
be defined from such data. It is also possible transform our
$N_{gals}$ to the MaxBCG $N_{200}$ using Eqn. 7, and plot the relation
from \citet{ryk08} using this pseudo-$N_{200}$; this is shown as the
dashed line in Figure~\ref{ngallx}. We caution that this is not a
reliable conversion because the richness transformation is difficult
and \citet{ryk08} use a completely different prescription for
computing $L_X$.

\begin{deluxetable*}{lrrrrrr}
\tabletypesize{\small}
\tablecolumns{7} 
\tablewidth{0pc} 
\tablecaption{X-ray Measurements}
\tablehead{
\colhead{}  & \multicolumn{3}{c}{Within $0.5 h^{-1}_{70}$ Mpc}  &  \multicolumn{3}{c}{Within $1.0 h^{-1}_{70}$ Mpc} \\
\cline{2-4} \cline{5-7} \\[-6pt]
\colhead{Name} & \colhead{$L_X$ ($10^{43}$ erg s$^{-1}$)} &  \colhead{Err($L_X$)} & \colhead{$T_X$ (keV)}  & \colhead{$L_X$ ($10^{43}$ erg s$^{-1}$)}  &  \colhead{Err($L_X$)} & \colhead{$T_X$ (keV)}
}
\startdata
NSC J111750+685910 &    0.620  & 0.140 & 0.8 &  0.160: & 0.260  & 0.4 \\ 
NSC J162305+653454 &    0.200: & 0.960  & 0.4 &         1.160  & 0.510 & 1.1 \\ 
NSC J091711+524442 &    2.120  & 0.600 & 1.4 &  28.760  & 0.220 & 5.3 \\ 
NSC J102307+520201 &    1.360  & 0.830 & 1.2 &  5.020  & 0.310 & 2.2 \\ 
NSC J101218+460643 &    1.780  & 0.520 & 1.3 &  1.020  & 0.380 & 1.0 \\ 
NSC J103805+420426 &    0.160  & 0.030 & 0.4 &  0.160: & 0.620  & 0.4 \\ 
NSC J173315+374215 &    0.360  & 2.450 & 0.6 &  5.160  & 1.320 & 2.3 \\ 
NSC J151120+363421 &    1.590  & 0.950 & 1.2 &  0.860  & 0.910 & 0.9 \\ 
NSC J082043+301238 &    0.120: & 0.000  & 0.3 &         1.050  & 0.160 & 1.0 \\ 
NSC J152111+292632 &    0.370  & 1.920 & 0.6 &  2.900  & 1.000 & 1.7 \\ 
NSC J081942+264129 &    0.020: & 0.090  & 0.2 &         3.190  & 0.040 & 1.8 \\ 
NSC J155312+273835 &    0.570  & 1.370 & 0.7 &  1.820  & 1.370 & 1.3 \\ 
NSC J020211+190446 &    5.700  & 0.250 & 2.4 &  8.510  & 0.290 & 2.9 \\ 
NSC J114047+181932 &    4.350  & 0.580 & 2.1 &  2.060  & 0.600 & 1.4 \\ 
NSC J164837+193606 &    0.270  & 0.610 & 0.5 &  0.280: & 0.530  & 0.5 \\ 
NSC J085246+161920 &    0.800  & 2.300 & 0.9 &  1.010: & 2.680  & 1.0 \\ 
NSC J141229+140110 &    4.480  & 0.630 & 2.1 &  6.560  & 0.520 & 2.5 \\ 
NSC J011144+100349 &    0.210  & 0.120 & 0.5 &  1.700  & 1.430 & 1.3 \\ 
NSC J094338+085430 &    0.190  & 2.010 & 0.4 &  1.920  & 1.540 & 1.4 \\ 
NSC J135224+092048 &    0.520  & 0.030 & 0.7 &  4.010  & 0.060 & 2.0 \\ 
NSC J021010+080844 &    0.690  & 1.450 & 0.8 &  5.850  & 1.370 & 2.4 \\ 
NSC J104929+033846 &    1.770  & 0.800 & 1.3 &  2.530  & 0.700 & 1.6 \\ 
NSC J154555+030814 &    0.930  & 0.030 & 1.0 &  0.640: & 0.260  & 0.8 \\ 
NSC J014426+021221 &    0.570  & 0.050 & 0.7 &  1.850  & 0.040 & 1.3 \\ 
NSC J104534-002506 &    0.180: & 0.460  & 0.4 &         0.450: & 0.050  & 0.7 \\ 
NSC J152156+013000 &    0.240: & 0.000  & 0.5 &         1.870  & 0.060 & 1.4 \\ 
\enddata 
\label{lx}
\end{deluxetable*} 

While the X-ray data from RASS is limited, especially for the poorer,
lower mass systems, this catalog of individual cluster X-ray
measurements is the largest compiled to date. It is only recently that astronomers have undertaken systematic comparisons of optical and X-ray cluster samples by returning to the source data and re-extracting physical properties consistently, rather than simply matching catalogs. For
instance,\citet{don02} compared independently detected X-ray and
optical clusters from the same patches of sky. They found poor
correlation between optical richness and X-ray luminosity, but could
not pinpoint the physical reason for this, and pointed out the need to
understand the effect of this scatter on mass selection. The RASS-SDSS
clusters survey \citep{pop04} instead uses a small but very well
measured sample of 114 X-ray detected clusters, and finds good
correlation between X-ray luminosity or temperature and optical
luminosity, if one has excellent data and chooses the measurement
parameters (such as the aperture for richness measurement)
carefully. However, it is worth noting that their sample remains one
requiring X-ray detections, which was shown by \citet{don02} to
potentially bias the results.

The only other large optical - X-ray comparisons are those of
\citet{dai07}, who used stacking techniques to derive X-ray properties
of over 4000 clusters selected optically from the Two-Micron All Sky
Survey (2MASS), and \citet{ryk08}, who stacked X-ray data for
$\sim17,000$ MaxBCG clusters from the SDSS. Even with the low redshift
limit ($z<0.1$) imposed by the shallow depth of 2MASS, \citet{dai07}
relied on stacking of X-ray data for clusters binned by their optical
properties to measure correlations between mass (optical richness),
luminosity and temperature. They find similar correlations to those in
the literature for individual clusters, but must model the Poisson
fluctuations in the number of galaxies in a cluster of a given
mass. At higher redshifts, where evolution in the cluster populations
becomes more important, understanding and modeling these fluctuations
will be more challenging. \citet{ryk08} were able to examine some
issues related to bias arising from scatter in the $L_X$-richness
relation with a small sample of clusters where individual X-ray
measurements were possible.

\section{Conclusions}

We have presented NoSOCS, a new cluster catalog based on the
$|b|>30^{\circ}$ plate scans from the Digitized Second Palomar
Observatory Survey. Spanning over $\pi$ steradians, this is the
largest area optical cluster catalog created since those of
\citet{abe58} and \citet{aco89}. In terms of area coverage, it will
only be superseded by new sky surveys such as Pan-STARRS \citep{kai04}
and LSST \citep{tys06}, both of which have cosmology through clusters
as important science drivers. We show consistency among the three
regions covered by NoSOCS, and with the SDSS MaxBCG cluster catalog of
\citet{koe07b}. However, interesting discrepancies between these two
large surveys remain. These include large numbers of poor clusters
missed by one survey but found in another, suggesting lower
completeness, higher contamination, or some combination of the two for
such systems, in either or both surveys. Even for supposedly rich
clusters there are sufficient discrepancies to call into question our
ability to use such surveys for high-precision cosmological
constraints. Understanding the sources of these disagreements
requires further investigation of both systematic errors and
individual cluster candidates. 

We have also derived X-ray luminosities for a large subset of our
cluster sample from ROSAT all-sky X-ray survey data. We demonstrate
that the optical richness and X-ray luminosities are well correlated,
albeit with moderate scatter. We find that, despite the poor
photometric data and low X-ray luminosities of most NoSOCS clusters,
the correlation between $L_{x}$ and $N_{gals}$ is in good agreement
with literature results using better data and stacking
analyses. Refinements to both the optical richnesses and especially
deeper X-ray survey data will be necessary to improve this relation
and truly understand the utility of optical richnesses for mass
estimation. Nevertheless, our results show promise for using large
surveys for such measurements in cases where the data quality is less
than superb, as may be expected for the highest redshift clusters even
in upcoming deep surveys such as Pan-STARRS and LSST. Furthermore, our
ability to measure X-ray luminosities for hundreds of clusters not
originally detected in the RASS argues for improved multi-wavelength
detection methods that leverage multiple surveys (optical, infrared,
X-ray, S-Z) to find distant and/or poor clusters which would otherwise
fall below the significance cutoff in a single passband.

\acknowledgments Processing and cataloging of DPOSS was supported by a
generous grant from the Norris Foundation, and by other private
donors.  SGD acknowledges partial support from the NSF grant
AST-0407448, and the Ajax Foundation. We would like to thank Dr. Peter
Kalberla for deriving the HI column densities at all of the cluster
positions. PAAL was supported by the Funda\c{c}\~{a}o de Amparo \`{a}
Pesquisa do Estado de S\~{a}o Paulo (FAPESP, processes 03/04110-3,
06/04955-1 and 07/04655-0). This work would have been impossible
without the DPOSS team,especially Stephen Odewahn, Robert Brunner and
Ashish Mahabal. The POSS-II photographic team at Palomar Observatory,
the STScI digitization team, numerous undergraduates who assisted with
the calibration.and the many people who worked on DPOSS all made
important contributions to the photometric survey.

\end{document}